\documentclass[12pt]{JHEP3}

\usepackage{amssymb}
\usepackage{amsmath}
\usepackage{graphicx}
\usepackage{latexsym}

\newcommand{\mhu}{m_{H_U}}
\newcommand{\mhd}{m_{H_D}}

\newcommand{\mer}{m_{{\tilde{e}_R}}}
\newcommand{\mel}{m_{{\tilde{e}_L}}}
\newcommand{\mmur}{m_{{\tilde{\mu}_R}}}
\newcommand{\mmul}{m_{{\tilde{\mu}_L}}}
\newcommand{\mslr}{m_{{\tilde{l}_R}}}
\newcommand{\msll}{m_{{\tilde{l}_L}}}
\newcommand{\msnu}{m_{{\tilde{\nu}}}}

\newcommand{\snu}{\tilde{\nu}_L}

\newcommand{\sql}{\tilde{q}_L}
\newcommand{\sul}{\tilde{u}_L}
\newcommand{\sdl}{\tilde{d}_L}
\newcommand{\sqr}{\tilde{q}_R}
\newcommand{\sur}{\tilde{u}_R}
\newcommand{\sdr}{\tilde{d}_R}

\newcommand{\staul}{\tilde{\tau}_1}

\newcommand{\sll}{\tilde{l}_L}
\newcommand{\slr}{\tilde{l}_R}
\newcommand{\mglu}{m_{{\tilde{g}}}}
\newcommand{\mupl}{m_{{\tilde{u}_L}}}
\newcommand{\mupr}{m_{{\tilde{u}_R}}}
\newcommand{\msql}{m_{{\tilde{q}_L}}}
\newcommand{\msqr}{m_{{\tilde{q}_R}}}

\newcommand{\mdnl}{m_{{\tilde{d}_L}}}
\newcommand{\mdnr}{m_{{\tilde{d}_R}}}
\newcommand{\mlsp}{m_{{\tilde{\chi}^0_1}}}
\newcommand{\mzii}{m_{{\tilde{\chi}^0_2}}}
\newcommand{\mziii}{m_{{\tilde{\chi}^0_3}}}
\newcommand{\mziv}{m_{{\tilde{\chi}^0_4}}}
\newcommand{\mtaul}{m_{\tilde{\tau}_1}}
\newcommand{\mtauh}{m_{\tilde{\tau}_2}}
\newcommand{\mntau}{m_{\tilde{\nu}_{\tau}}}
\newcommand{\mbl}{m_{\tilde{b}_1}}
\newcommand{\mbh}{m_{\tilde{b}_2}}
\newcommand{\mtl}{m_{\tilde{t}_1}}
\newcommand{\mth}{m_{\tilde{t}_2}}
\newcommand{\ziv}{\tilde{\chi}^0_4}
\newcommand{\ziii}{\tilde{\chi}^0_3}
\newcommand{\wiipm}{\tilde{\chi}^\pm_2}
\newcommand{\wipm}{\tilde{\chi}^\pm_1}
\newcommand{\wii}{\tilde{\chi}^+_2}
\newcommand{\wi}{\tilde{\chi}^+_1}
\newcommand{\wiic}{\tilde{\chi}^-_2}
\newcommand{\wic}{\tilde{\chi}^-_1}
\newcommand{\mwi}{m_{\tilde{\chi}^+_1}}
\newcommand{\mwii}{m_{\tilde{\chi}^+_2}}
\newcommand{\lsp}{\tilde{\chi}^0_1}
\newcommand{\zii}{\tilde{\chi}^0_2}
\newcommand{\sq}{\tilde{q}}
\newcommand{\psla}{p\kern-.45em/}
\newcommand{\esla}{E{\!\!\!/}}

\newcommand{\fb}{{\rm  fb}}

\newcommand{\non}{\nonumber}
\newcommand{\be}{\begin{equation}}
\newcommand{\ee}{\end{equation}}
\newcommand{\beq}{ \begin{eqnarray} }
\newcommand{\eeq}{ \end{eqnarray} }
\newcommand{\beqstar}{ \begin{eqnarray*} }
\newcommand{\eeqstar}{ \end{eqnarray*} }
\newcommand{\gsim}{ \mathop{}_{\textstyle \sim}^{\textstyle >} }

\newcommand{\GEV}{ {\rm GeV} }

\newcommand{\wtl}{\widetilde{L}}
\newcommand{\wte}{\tilde{e}}

\newcommand{\sml}{\tilde{\mu}_L}

\newcommand{\sel}{\tilde{e}_L}
\newcommand{\ser}{\tilde{e}_R}

\newcommand{\msl}{m_{\tilde{l}}}

\newcommand{\eps}{\epsilon}
\newcommand{\meg}{\mu\rar e\gamma}
\newcommand{\amu}{\delta a_\mu}
\newcommand{\amususy}{a_\mu^{\rm SUSY}}
\newcommand{\rar}{\rightarrow}

\title{Discriminating Electroweak-ino Parameter Ordering at the LHC and Its Impact on LFV Studies}

\author{Junji Hisano\\
Institute for Cosmic Ray Research (ICRR),\\
University of Tokyo, Kashiwa, Chiba 277-8582, Japan\\
IPMU, Tokyo University, Kashiwa, Chiba, 277-8568, Japan}

\author{Mihoko M. Nojiri\\
IPMU, Tokyo University, Kashiwa, Chiba, 277-8568, Japan\\
Theory Group, KEK,1-1 Oho, Tsukuba, Ibaraki 305-0801, Japan\\
The Graduate University for Advanced Studies (Sokendai), \\
1-1 Oho, Tsukuba, Ibaraki 305-0801, Japan}
\author{Warintorn Sreethawong\\
Theory Group, KEK,1-1 Oho, Tsukuba, Ibaraki 305-0801, Japan\\
The Graduate University for Advanced Studies (Sokendai), \\
1-1 Oho, Tsukuba, Ibaraki 305-0801, Japan}

\preprint{KEK-TH-1302}

\abstract{Current limit on the dark matter relic abundance may suggest that $|\mu|$ should be smaller than prediction in the minimal supergravity scenario (mSUGRA) for moderate $m_0$ and $m_{1/2}$. The electroweak-ino parameter $M_1, M_2$ and $|\mu|$ are then much closer to each other. This can be realized naturally in the non-universal Higgs mass model (NUHM). Since the heaviest neutralino ($\ziv$) and chargino ($\wiipm$) have significant gaugino components, they may appear frequently in the left-handed squark decay and then be detectable at the LHC. In such a case, we showed that the hierarchy of $M_1, M_2$ and $|\mu|$ can be determined. 
In the light slepton mass scenario with non-vanishing lepton-flavor violation (LFV) in the right-handed sector, NUHM with small $|\mu|$ corresponds to region of parameter space where strong cancellation among leading contributions to $Br(\meg)$ can occur. We showed that determination of electroweak-ino hierarchy plays a crucial role in resolving cancellation point of $Br(\meg)$ and determination of LFV parameters. We also discussed test of the universality of the slepton masses at the LHC and the implications to SUSY flavor models.
}

\begin{document}
\section{Introduction}

Even though the Standard Model (SM) describes the Nature very well, there are a number of theoretical and phenomenological issues that the SM cannot give a plausible explanation. The notorious one is the hierarchy problem. In the absence of a symmetry to protect the mass of the Higgs boson from the radiative corrections, its natural value should be of order of UV cutoff scale. However, from the precision electroweak measurements, the SM Higgs mass is lower than $193~\GEV$ at the $95 \%$ confidence level \cite{LEP}.
On the cosmological side, while the evidence of the dark matter (DM) in the universe is now available from a variety of observational data \cite{wmap_new}\cite{sdss_new}\cite{wmap}\cite{IaSN}\cite{sdss}\cite{Clowe}, the SM cannot provide a viable candidate for it. 
Therefore, it is convincing that the SM must be viewed as a low energy effective theory of the more fundamental theory.

Supersymmetry (SUSY) is one among the most promising candidates of the physics beyond the SM. It provides a natural solution to the hierarchy problem and allows the unification of gauge couplings in the context of grand unified theory (GUT). Moreover, in the model with conserved R-parity, the lightest supersymmetric particle (LSP), typically the lightest neutralino, is stable and provides a dark matter candidate.

In general SUSY models, the non-vanishing off-diagonal elements of the slepton mass violate the lepton-flavor conservation and then are strongly constrained by experiments \cite{constraint lfv}. Since arbitrary sfermion masses generally lead to flavor-changing-neutral-current (FCNC) processes exceeding the experimental bounds, universality among generation of sfermion masses is usually imposed at some high energy scale. The most popular one is the minimal supergravity scenario (mSUGRA) \cite{msugra},
whose input parameters are the universal scalar mass $m_0$, 
the universal trilinear scalar coupling $A_0$, the gaugino mass $m_{1/2}$ at the GUT scale and $\tan\beta$. 

Only small regions of mSUGRA parameter space have DM relic density within the observational bound \cite{dm_msugra}. This is because $|\mu|\gg m_{1/2}$ is predicted for moderate $m_0 \sim m_{1/2}$ and the LSP is almost pure Bino in this scenario. For moderate $\tan\beta$, only $t$-channel $\slr$ exchange diagram gives important contribution to DM annihilation cross section \cite{dm_ffbar}. Therefore, the cross section is typically too small and the relic abundance exceeds the experimental upper bound. However, when relaxing the universality condition of the SUSY breaking parameters at the GUT scale, one can freely adjust the thermal relic density.

One of the well-motivated relaxations is to allow soft scalar Higgs masses to be varied such as in the non-universal Higgs mass model (NUHM) \cite{nuhm1}\cite{nuhm2}\cite{Nojiri_D63}\cite{Hisano_D65}. 

Supersymmetric particles may be discovered in the early stages of LHC data collection. The squarks and gluino with masses below 1.5 TeV are expected to be found at the LHC for ${\cal L } =1~\rm{fb}^{-1}$ at  $\sqrt{s}=14$~TeV \cite{atlas and cms}. 
By studying the kinematics of long cascade decays, 
one can extract information on masses of sparticles 
involved and more or less constrain the parameter space of the underlying 
theory. 
 This is especially successful when sleptons are 
involved in the cascade decays. However, 
it was pointed out that, for a given set of experimental SUSY signatures at the LHC, more than one set of parameters could be present and ambiguities predominantly occur in the electroweak-ino sector. This is sometimes called the LHC Inverse Problem \cite{Arkani-Hamed}. A similar result was obtained in \cite{Lafaye} when an exclusive likelihood map of SPS1a point is studied by a weighted Markov chain technique. Even for the favorable SPS1a point, we have multiple solutions. 

In this paper, we study the interplay between the LHC measurements of the  
parameters in electroweak (EW)  sector and lepton-flavor violation. 
The LFV processes are important discovery channels of physics beyond the standard model.  Right now, the main constraints come from the upper bounds of 
rare decay process searches \cite{mega}\cite{belle}\cite{babar}:
\begin{eqnarray}
Br(\mu\rar e\gamma)&<& 1.2\times 10^{-11},\\
Br(\tau\rar \mu\gamma)&<& 4.5(6.8)\times 10^{-8},\\
Br(\tau\rar e\gamma)&<& 1.2(1.1)\times 10^{-7}.
\end{eqnarray}
The bounds for tau LFV processes come from Belle (Babar). On-going MEG experiment \cite{meg} aims to push the sensitivity of $\meg$ down by two orders of magnitude, and the SuperKEKB \cite{superkekb} and SuperB Factory \cite{superb} aim to improve the sensitivity of the $\tau$ decay by one or two orders of magnitude in near future. 

If slepton is directly observed at the LHC, together with these advanced LFV measurements, we will be able to study the flavor sector of supersymmetry in detail. Several papers \cite{Hisano_D65}\cite{Allanach_D77}\cite{Hirsch_D78} recently point out that the ability of the LHC to measure the slepton mass differences will significantly constrain the possible scenarios. The further improvement may be obtained through the constraints on $U(1)$ and $SU(2)$ gaugino masses, $M_1$ and $M_2$, and $\mu$ parameter from the neutralino and chargino mass measurements at the LHC. Note  that the inverse  problem has an  impact on the $\meg$ study. As will be discussed in detail later, the $Br(\meg)$ could be suppressed due to cancellation among different diagrams \cite{Hisano_D53}. The severity of the cancellation depends strongly on value of $\mu$ 
parameter and slepton masses. Accordingly, to unravel the off-diagonal elements from the measurement of rare decay processes such as  $Br(\meg)$, one needs to solve the inverse problem and determine the SUSY parameters precisely. 

In this paper, we take the non-universal Higgs mass model with positive $\mu$ with $\mu \sim m_{1/2}$ due to the DM relic abundance reason. In addition, 
we choose a mass hierarchy $\mlsp <\mer <\mzii <\mel <\mziv$ 
so that both left- and right-handed sleptons appear in neutralino decay. 
Notably, we find  
$\mer<\mzii$ and 
$\mel<\mziv$ should be  satisfied 
 when there is strong cancellation among the 
LFV diagrams involving the right-handed sleptons for the 
model points  with correct DM relic abundance in the NUHM.  
For this parameters,  many cascade decay modes 
 involving sleptons can be  observed at the LHC. 
Using the information, a solution of the EW parameters 
can be selected among the 
 multiple solutions, and ambiguity in the MSSM parameters 
would be reduced significantly, leading better prediction to 
$Br(\mu \rightarrow e \gamma)$ from the LFV slepton masses. 
 
The paper is organized as follows. 
In the next section, we introduce models we study.
The sparticle  mass determination is studied by explicit 
MC simulation in Section III for our sample model point.  
In Section IV, we discuss the ambiguity of the corresponding MSSM parameters 
 and demonstrate that it can be removed by studying decay branching ratios 
 of $\tilde\chi_i\rar 2l+X$, charge asymmetry of the decay distribution, and the rate into 2 hard jets + missing $E_T$ mode at the LHC.
The implication to LFV studies is discussed in Section V.
We emphasize the importance of the slepton masses determination 
and compare the sensitivity to the off-diagonal elements of 
 slepton mass matrices at the 
LHC and the other LFV searches such as MEG experiments. 
We also study the relation among slepton masses and $\mu$ parameter 
at the point of $Br(\mu \rightarrow e \gamma)$ cancellation. We
show that the MSSM parameters would  be studied very precisely 
at the LHC  through the cascade decays involving sleptons. 
 Finally, Section VI is devoted to discussions.

\section{Models}

The models under consideration in this paper consist of two important features.
The first one is related to the scale of sparticle masses and their hierarchy. We chose the NUHM model as a representative. The second class involves issue of lepton-flavor-violation. Here, two models are selected: the MSSM with right-handed neutrinos and the MSSM with horizontal symmetry. Although models of interest have both NUHM and LFV, the detail descriptions of these features are given separately below.

\subsection{Non-universal Higgs Mass Model}

This model is motivated by the supersymmetric grand unified theories in which the Higgs fields do not belong to the same multiplet as the matter fields. Therefore, it is natural to expect that different multiplets would have different soft masses at the GUT scale. 

In mSUGRA scenario, for the correct pattern of EWSB, $m^2_{H_U} + \mu^2 > -m_Z^2/2$ must be satisfied at weak scale. Since the SUSY breaking mass squared for the Higgs doublet $H_U$, $\mhu^2$, is always driven to large negative value at weak scale by large top Yukawa coupling, $|\mu|$ is typically large. A smaller $|\mu|$ value can be easily obtained if one relaxes universality condition, especially $\mhu > m_0$ at the GUT scale. Once $|\mu|\sim m_{1/2}$, the LSP has larger Higgsino components. In that case, $s$-channel $Z/h^0$ exchange contributions and $\lsp\lsp\rar WW$ contribution to the DM pair-annihilation cross section become more important \cite{dm_ffbar}\cite{dm_zh} and the relic abundance would be small to be compatible with observational value. 

On the collider physics side, the reduction of $|\mu|$ would lead to a rich pattern of  colored sparticle cascade decays at the LHC. Due to their considerable Wino  components, $\ziv$ and $\wiipm$ can be produced copiously in $SU(2)$ doublet squark decay. Their successive cascade decays are the followings \cite{Nojiri_D63}:
\beq\label{heavier_decay}
	\wii &\rar&\tilde\nu_L\rar\wi  \,,\non\\
	\ziv &\rar&\slr\rar\lsp \,,\\
	\ziv &\rar&\sll\rar\zii, \lsp \,,\non
\eeq
which lead to clean, fruitful opposite sign same flavor (OSSF) dilepton events. The maximum values of dilepton invariant mass distributions depend on the cascade decay chains. If several endpoints can be identified and measured, it gives access to sparticle masses, or at least, their mass differences.

\subsection{LFV Models}

In the MSSM the low-energy LFV processes are induced by the
off-diagonal terms in the slepton mass matrices, which depend on the
origin of the SUSY breaking and physics beyond the MSSM.  In this
paper we discuss two models which predict LFV, i) MSSM with
right-handed neutrinos and ii) MSSM with horizontal symmetries.

Before going to those models, we first discuss $\meg$ process
and an anomalous magnetic dipole moment $a_{l_j}\equiv (g-2)_{l_j}$.
The effective operators for $l^-_j\rar l^-_i\gamma$ and $a_{l_j}$ are
written as
\beq
	{\mathcal L}_{eff}&=& e\frac{m_{l_j}}{2}~\bar{l_j}\sigma^{\mu\nu}F_{\mu\nu}
	(A^L_{ij}P_L+A^R_{ij}P_R)l_i\label{operator}
\eeq
where $m_{l_j}$ is a mass of the charged lepton $l_j$. The coefficients $A^L_{ij}$ and $A^R_{ij}$ are functions of masses and mixings of sparticles inside the loop corrections. Their full formula can be found in \cite{Hisano_D53}.
The SUSY contributions to the anomalous magnetic dipole moment of the muon, $\amususy$, is given by
\beq
	\amususy &=&m^2_\mu (A^L_{22}+A^R_{22}).
\eeq
The branching ratio for the decay $l^-_j\rar l^-_i\gamma$ is
\beq
	Br(l^-_j\rar l^-_i\gamma)&=&\frac{48\pi^3\alpha}{G^2_F}(|A^L_{ij}|^2+|A^R_{ij}|^2) \times
Br(l^-_j\rar l^-_i \bar{\nu}_i\nu_j).
\eeq

The experimental value of the anomalous magnetic moment of the muon measured by the E821 experiment at Brookhaven with extremely high precision \cite{E821} is given by
\beq
	a_\mu^{\textrm{exp}}&=&116592080(63)\times 10^{-11}.
\eeq
According to the most recent calculations of the hadronic contribution based on the $e^+e^-$ data \cite{amu_HLO}, the difference between the SM prediction and the experimental value is 
\beq
	\amu =a_\mu^{\textrm{exp}} - a_\mu^{\textrm{SM}}
	&=& +302(88)\times 10^{-11},\label{amu}
\eeq
which corresponds to 3.4-$\sigma$ deviations. It was pointed out that this discrepancy tends to come from new physics contributions as it is unlikely to be explained by errors in the determination of the hadronic contributions \cite{Passera}. Moreover, in the context of SUSY, the $\amu$ anomaly, Eq.~(\ref{amu}), suggests light slepton-chargino sector. Accordingly, unless flavor mixings are very small, $\meg$ may be detected at MEG experiment \cite{meg} soon.
  
We now explain two LFV models and discuss their phenomenology.  

\subsubsection{MSSM with Right-handed Neutrinos}

This model is motivated by the observations of non-vanishing neutrino masses and neutrino oscillation. In the model with three right-handed neutrinos, the superpotential for the lepton sector is given by
\beq
	W &=& f_l^{ij}E^c_i L_j H_D+ f_\nu^{ij} N^c_iL_jH_U
	+\frac{1}{2}M_\nu^{ij}N^c_i N^c_j,
\eeq
where $L_i, E^c_i$ and $N^c_i$ represent chiral multiplets for
left-handed lepton doublet, right-handed charged lepton singlet and
right-handed neutrino singlet respectively, and $H_U$ and $H_D$ for
two Higgs doublets with opposite hypercharge. The neutrino mass matrix
can be obtained by the seesaw mechanism and is given by
\beq
	m_\nu &=& f^T_\nu M_\nu^{-1} f_\nu \langle h_U\rangle^2
\eeq
where $\langle h_U\rangle$ is the vacuum expectation value of the
neutral Higgs component $h_U$ of the $H_U$ multiplet.

In general, the Yukawa couplings $f_l$ and $f_\nu$ cannot be
diagonalized simultaneously. Therefore, even though one assumes that
mass matrices for left-handed and right-handed sleptons,
  $m^2_{\wtl}$ and $m^2_{\wte}$, are proportional to the unit matrix
at the GUT scale, the LFV masses of $m^2_{\wtl}$ will always be
generated via the renormalization group evolution,
\beq
	\mu\frac{d}{d\mu}(m^2_{\wtl})^j_i &=&
		\left(\mu\frac{d}{d\mu}(m^2_{\wtl})^j_i\right)_{\textrm{MSSM}}
		+\frac{1}{16\pi^2}\left[ (m^2_{\wtl}f^\dagger_\nu f_\nu 
		+ f^\dagger_\nu f_\nu m^2_{\wtl})^j_i \right.\non\\
		&& \left. + 2(f^\dagger_\nu m^2_{\wtl} f_\nu 
		+ \tilde m^2_{h_2}f^\dagger_\nu f_\nu + A^\dagger_\nu A_\nu )^j_i\right],
\eeq
by the non-vanishing off-diagonal elements of $f_\nu$ and of the
trilinear soft parameter $A_\nu$ \cite{Borzumati:1986qx}. The large mixing angles observed in the atmospheric and solar neutrino oscillation experiments enhance the
LFV masses of $m^2_{\wtl}$, and the experimental bounds on $Br(\meg)$ and
$Br(\tau\rightarrow \mu\gamma)$ give stringent constraints on the model \cite{Hisano:1998fj}.

It has been studied that if the LFV is generated only in the
left-handed slepton sector, $Br(\meg)$ will be strongly correlated with
$(\amususy)^2$ \cite{Hisano510}. This is because the
chargino-sneutrino diagram dominates over other contributions in both
observables. By taking a common mass for all sparticles, the current
experimental bounds on $\amususy(\equiv\amu)$ and $Br(\meg)$ put a
stringent constraint on the left-handed LFV slepton mass term,
\beq
	\left|\frac{(m_{\tilde L}^2)^{\mu}_e}{m^2_{\rm SUSY}}\right| &\lesssim&
	2\times 10^{-4}\left(\frac{|\amususy|}{3.02\times 10^{-9}}\right)^{-1}
	\left(\frac{Br(\meg)}{1.2\times 10^{-11}}\right)^{\frac{1}{2}}.
\eeq

On the other hand, we still have a room for sizable mass difference
between the left-handed selectron and other sleptons, which may be
measured at the LHC. Without LFV, basically the ``left-handed'' tau
  slepton mass is deviated from other left-handed slepton masses
  due to its Yukawa coupling effect to the renormalization group
  evolution or the left-right mixing term, while mass difference
$m_{\sel}-m_{\sml}$ is negligibly small. However, in the case of
nonzero LFV, mass splittings could be larger and would be detectable
at the LHC.

\subsubsection{MSSM with Horizontal Symmetry}

Horizontal symmetries are introduced to derive the hierarchical
structure in the Yukawa coupling constants \cite{Froggatt} and also to suppress
the off-diagonal terms in the sfermion mass matrices \cite{Leurer_B420}\cite{Grossman_B448}. 

In this paper, we studied the MSSM with horizontal symmetry, which is
originated from one of models in \cite{Feng}.  The original model was
a supersymmetric model in which the SUSY breaking is mediated to
  the MSSM by gravity interactions. The
gravity-mediated contribution is assumed to be subject to an
approximate horizontal symmetry. More specifically, consider a
horizontal $U(1)\times U(1)$ symmetry where each $U(1)$ is explicitly
broken by a scalar singlet spurion carrying the corresponding charge
$-1$. The sizes of both $U(1)$ breakings are assumed to be equal and then
parametrized by a single parameter $\eps\sim |V_{us}|\sim 0.2$. The
horizontal charge assignment for lepton sector is as follows:
\beq
	L_1(4,0),~~L_2(2,2),~~L_3(0,4); \non\\
	\bar{E}_1(1,0),~~ \bar{E}_2(1,-2),~~ \bar{E}_3(0,-3).
\eeq

Using above horizontal symmetry, we parameterize the left- and right-handed slepton mass matrices at the GUT scale as
\beq
	m^2_{\wtl}=m^2_0+x m_0^2 X'_{\wtl},~~~~~
	M^2_{\wte}=m^2_0+xm^2_0 X'_{\wte},
\eeq
where $x$ is the ratio between the flavor-independent and dependent contributions. The structure of matrices $X'_L$ and $X'_R$ can be determined uniquely by selection rules by the horizontal symmetry and have the following forms:
\beq
	X'_{\wtl}\sim\left(\begin{array}{ccc}
			0&\eps^4&\eps^8\\ \eps^4&0&\eps^4\\ \eps^8&\eps^4&0
			\end{array}\right),~~~~~
	X'_{\wte}\sim\left(\begin{array}{ccc}
			0&\eps^2&\eps^4\\ \eps^2&0&\eps^2\\ \eps^4&\eps^2&0
			\end{array}\right).
\eeq
Here, we neglect the flavor-dependent contribution to the
flavor-diagonal mass terms, though they are also allowed
by the symmetry with size $\sim x m_0^2$.
It is argued in \cite{Feng} that even $x\gsim 1$ is allowed from
phenomenological constraints under the horizontal symmetry when
$m_0<380$~GeV, $m_{1/2}<160$~GeV, and $5<\tan\beta<15$. 

To discuss the $\meg$ phenomenology of this model, first of all, it should be remembered that the effective $\meg$ operator, Eq.~(\ref{operator}), must flip the chirality and change the flavor of the external leptons. Since the left-handed LFV masses are suppressed in this model, the dominant LFV contribution comes from the right-handed sector. In the mass insertion approximation, there are two dominant one-loop diagrams when $M_1\mu\tan\beta$ is large \cite{Hisano391}. The former one has chirality flip on an internal line via left-right slepton mixing and its contribution is 
\beq
	{A^L_{12}|}_{\widetilde B^0} &=& 
		\frac{1}{2}\frac{\alpha_Y}{4\pi}\frac{1}{\overline m^2_{\slr}}
		\frac{(m^2_{\slr})^\mu_e}{\overline m^2_{\slr}}
		\frac{M_1(A_\mu+\mu\tan\beta)}{\overline m^2_{\sll}}.
\eeq

In the latter diagram, the chirality is flipped at one vertex via Yukawa coupling. 
The contribution is proportional to the Higgsino components,
which come from the Bino-Higgsino mixing of neutralino.
The amplitude is given by
\beq
	{A^L_{12}|}_{(\widetilde{B}^0 -\widetilde{H}^0 \rm{mixing})} &=& 
		-\frac{\alpha_Y}{4\pi}\frac{1}{\overline m^2_{\slr}}
		\frac{(m^2_{\slr})^\mu_e}{\overline m^2_{\slr}}
		\frac{M_1\mu\tan\beta}{\overline m^2_{\slr}}
		f_1(\frac{\mu^2}{\overline m^2_{\slr}}).
\eeq
In the above equations, $\overline m^2_{\slr}(\overline m^2_{\sll})$ stands for an averaged right-handed (left-handed) slepton mass, and $(m^2_{\slr})^\mu_e$ is a $(\mu,e)$ component of the right-handed charged slepton mass squared matrix. The kinematic function $f_1$ is given by
\beq
	f_1(x)&=&-\frac{8 - 11x + 4x^2 -x^3 + 2(2+x)\log x}{2(1-x)^4}
\eeq
which is a positive-definite, decreasing function of $x$.

A key point is that the relative sign of the two amplitudes is negative and then cancellation between diagrams can occur significantly. For $A_\mu =0$, a severe cancellation happens when
\beq
	\frac{1}{2\overline m^2_{\sll}}-\frac{1}{\overline m^2_{\slr}}f_1(\frac{\mu^2}{\overline m^2_{\slr}})~\sim~0.\label{cancellation_cond}
\eeq
This cancellation occurs when $\overline m^2_{\sll}\sim\mu^2$ in the limit of $\mu^2/{\overline m^2_{\slr}}\gg 1$.
Since dominant contributions for $\meg$ and $\amususy$ are now
different, then their correlation becomes much weaker than in the case
that the left-handed sleptons have  LFV masses.

Next, we consider the processes $\mu\rar eee$ and $\mu\rar e$
conversion in the nuclei.  Their formulae are also given in
  \cite{Hisano_D53}. See also Ref.~\cite{Kitano:2002mt} for
    precise evaluation of $\mu\rar e$ conversion in the nuclei. These
  two observables also suffer from the partial cancellation
  \cite{Hisano391}. The $\mu\rar eee$ receives dominant contribution
  from the penguin-type diagrams which are enhanced at large
  $\tan\beta$ region in a similar manner to $\meg$ so that their
  behaviors are alike; the subdominant box-type contribution just
  helps lifting up the depth of the cancellation valley. On the other
  hand, the $\mu\rar e$ conversion rate behaves rather differently and
  plays a complementary role in resolving the cancellation
  point. Thus, the correlation among the LFV processes are also weaker
  than the case that the left-handed sleptons have LFV masses.

\section{Monte Carlo Study}

In this section, we study the leptonic SUSY signals at the LHC in the non-universal Higgs mass scenario. For our analysis, we use ISAJET v7.75 \cite{isajet} to calculate the sparticle spectrum and IsaReD \cite{isared}, which is part of the IsaTools package, to evaluate the dark matter relic density. We generated $ 5 \times 10^6$ events by HERWIG 6.5 \cite{herwig}; this corresponds to about 300 $\fb^{-1}$ of integrated luminosity. The AcerDet package \cite{acerdet} is employed to simulate the detector response. 

\subsection{Model Point}
As mentioned earlier, we chose $\mhd = m_0 \neq \mhu$ at the GUT scale. We are interested in the case that both left- and right-handed sleptons can be directly produced via neutralino and chargino cascade decays. Then slepton masses should be light and a relatively small value of $m_0$ had been chosen. Furthermore, we chose a moderate value for $m_{1/2}$. The relevant parameters and sparticle masses for our studied point A are listed in Table \ref{pta}. 
We took $m_0=100 ~\GEV, m_{1/2}=300\GEV, \mhu=380~\GEV$. This leads to $\mu=271~\GEV, \mel=240~\GEV$, and $\mer=130~\GEV$. We will see that this corresponds to the point where strong cancellation among contributions to $Br(\meg)$ occurs.

\TABLE[!ht]{
\begin{tabular}{|c|c||c|c||c|c||c|c|}\hline
 $m_0$  & 100  & $m_{1/2}$ & 300  &  $\mhd$ & 100 &$\mhu$ & 380 \\\hline
$\tan\beta$ & 10 &  $\mu$& 271.33 & $M_1$ & 122.49  &  $M_2$& 230.89\\ \hline
\hline
$m_{\tilde{g}}$& 719.67 & & & & & & 
\\ \hline
$\mupl$ & 665.19 & $\mupr$ & 648.85 & 
$\mdnl$ & 670.29  &$\mdnr$ & 642.47
\\ \hline
$\mbl$ & 600.91 & $\mbh$ & 638.72 &
$\mtl$ & 462.35 & $\mth$ & 655.20
\\ \hline
$\mel$ & 239.62  & $\mer$ & 130.38  & 
$\mtaul$ & 128.07 & $\mtauh$ & 238.89
\\ \hline 
$\msnu$ &  224.37 &$\mntau$ & 222.16 &  
$\mwi$ & 196.30 & $\mwii$ & 321.62
\\ \hline
$\mlsp$ & 114.70 & $\mzii$ & 197.82  & 
 $\mziii$ & 278.87  &$\mziv$ & 323.23
 \\ \hline
  $m_h$ & 111.22 & $m_{H^0}$& 350.05 &  
  $m_A$ & 347.31 &  $m_{H^+}$& 358.56
  \\ \hline
\end{tabular}
\caption{\footnotesize Relevant parameters and sparticle masses in GeV for the parameter point A.} \label{pta}
}

\TABLE[!ht]{
\begin{tabular}{|c|c|c||c|c|c|}
\hline
& ~~~~~A~~~~~& mSUGRA  & & ~~~~~A~~~~~ & mSUGRA \\ \hline
$\sul\rar\zii$ & 25.4 & 31.6 & $\ziv\rar\sel$ & 2.3 & 1.0\\ \hline  
$\sul\rar\ziii$ & 0.2 & 0.1 & $\ziv\rar\ser$ & 0.7 & 0.3\\ \hline 
$\sul\rar\ziv$ & 7.8 & 1.3 & $\zii\rar\ser$ & 13.5 & 3.1\\ \hline
$\sul\rar\wi$ & 53.2 & 64.4 & $\wii\rar\snu$ & 7.4 & 2.1\\ \hline
$\sul\rar\wii$ &13.1& 1.8 & $\wi\rar\staul$ & 88.8 & 63.3\\ \hline 
$\sdl\rar\zii$& 21.8 & 30.7 & $\snu\rar\wi$ & 26.1 & -\\ \hline 
$\sdl\rar\ziii$& 0.4 & 0.1 &  $\sel\rar\zii$ & 26.5 & 5.9\\ \hline
$\sdl\rar\ziv$& 9.7 & 1.8 &  $\sel\rar\lsp$ & 35.9 & 84.0\\ \hline
$\sdl\rar\wic$& 42.1 & 60.4 & $\ser\rar\lsp$ & 100 & 100 \\ \hline
$\sdl\rar\wiic$& 23.1& 4.8 & && \\ \hline
\end{tabular}
\caption{\footnotesize Relevant branching ratios in \% for our sample 
point A compared with mSUGRA point.}\label{br_pta}
}

For this choice of parameters, the value of $\mu$ substantially reduces from $\mu=397.30 ~\GEV$ for mSUGRA case with inputs: $m_0=100~\GEV, m_{1/2}=300~\GEV,$ and $\tan\beta=10$. This increases the Higgsino components of the LSP as the mixing matrix elements $(N_{\widetilde B}, N_{\widetilde W}, N_{\widetilde H_D}, N_{\widetilde H_U}) = ( -0.96, 0.09, 0.22, -0.11)$. The dark matter relic density is evaluated to be 0.1179 which is consistent with the combined results from WMAP and SDSS \cite{wmap_new}\cite{sdss_new}, $\Omega_{\rm DM}h^2 = 0.111^{+0.011}_{-0.015}~(2\sigma)$. 

The relevant sparticle decay branching ratios are compared with mSUGRA case in Table \ref{br_pta}. In mSUGRA scenario, $\lsp$ is almost pure Bino, and $\zii$ and $\wipm$ are Wino-like. Then $\sql$ decays substantially into $\zii$ and $\wipm$. As the decay into $\sel$ is kinematically forbidden for this parameter choice, they essentially decay into left-handed component of $\staul$ with large branching and small Bino component of $\zii$ could decay into $\ser$ producing the well-known edge of dilepton invariant mass distribution \cite{Hinchliffe_D55}. On the other hand, heavier inos are Higgsino-like and would not be produced at the LHC. This will then forbid $\sel$ to show up at the LHC also.

For point A, when $\mu$ is smaller, heavier inos have larger Wino components so that their productions in the left-handed squark decay become significant. Table \ref{br_pta} shows that $Br(\sql\rar\ziv/\wii)$, along with $Br(\ziv\rar\sel)$ and $Br(\wii\rar\snu)$, are enhanced considerably. On the contrary, the enhancement in $Br(\zii\rar\ser)$ well demonstrates the increase of Bino component in $\zii$. Summarily, with small $\mu\sim m_{1/2}$, a number of sparticles could show up through various decay patterns, Eq.~(\ref{heavier_decay}), at the LHC experiment. Note that $\ziii$ is almost pure Higgsino and then is not produced neither at point A nor mSUGRA point.

\subsection{Two-lepton channel}
In analyzing SUSY signals at the LHC, we put an emphasis on lepton channels
when lepton presumably means electron or muon.
In this section, we focus on the celebrated dilepton invariant mass distribution and model independent constraints on the sparticle masses. The other signatures will be discussed in the next section. The dominant neutralino and chargino cascade decay processes which lead to OSSF lepton pair in the final states and the corresponding expected kinematics endpoints are listed in Table \ref{endp}
\TABLE[!ht]{
\begin{tabular}{|l||c|c|c|c|c|}
\hline
decay mode & $m^{\rm max}_{jll}$ & $m^{\rm min}_{jll}$& 
$m^{\rm max}_{jl}$ & $m^{\rm min}_{jl}$ & 
$m_{ll}^{\rm max}$  
\\ \hline
\hline
(1)\ $\sql\rar\zii\rar\ser\rar\lsp$
& 517.4 & 209.1 & 477.6 & 272.7 & 70.7
\\ \hline
(2)\ $\sql\rar\wii\rar\snu\rar\wi$ 
& 461.2 &  202.8 & 417.2 & 253.8 &111.6
\\ \hline
(3)\ $\sql\rar\ziv\rar\sel\rar\zii$
& 459.8 & 223.4 & 390.2 & 285.7 & 122.4
\\ \hline
(4)\ $\sql\rar\ziv\rar\ser\rar\lsp$
& 550.3 & 217.1 & 532.0 & 249.6 & 140.6
\\ \hline
(5)\  $\sql\rar\ziv\rar\sel\rar\lsp$
& 550.5 & 296.5 & 510.4 & 383.6 & 190.5
\\ \hline
\end{tabular}
\caption{\footnotesize Endpoints of invariant mass
distributions in GeV for various decay processes.}\label{endp}
}

The events are selected by the 
following criteria \cite{Bachacou_D62}:
\begin{itemize}
\item
an OSSF dilepton pair where both leptons have $p^l_T > 10 ~\GEV$ and $|\eta| < 2.5$,
\item
more than 4 jets with $p^j_{T,1} > 100 ~\GEV, p^j_{T,2,3,4} > 50 ~\GEV$,
\item
$M_{\textrm{eff}} \equiv p_{T,1} + p_{T,2} + p_{T,3} + p_{T,4} + \esla_T >  400 ~\GEV$,
\item
$\esla_T >$ max$(100, 0.2M_{\textrm{eff}})$.
\end{itemize}

The dilepton invariant mass distributions are shown in Figure \ref{mll}. In these plots, 
the subtraction of opposite sign opposite flavor (OSOF) dilepton distribution is utilized to reduce SUSY backgrounds. The endpoint position from $\sql\rar\tilde\chi_i\rar\tilde l\rar\tilde\chi_j$ is given by analytical formula \cite{Hinchliffe_D55}
\beq
	m_{ll}^{\rm max}&=& 
	\sqrt{\frac{(m^2_{\tilde\chi_i}-m^2_{\tilde l})(m^2_{\tilde l}-m^2_{\tilde\chi_j})}
	{m^2_{\tilde l}}}
\eeq

\FIGURE[!ht]{
\includegraphics[scale=0.3]{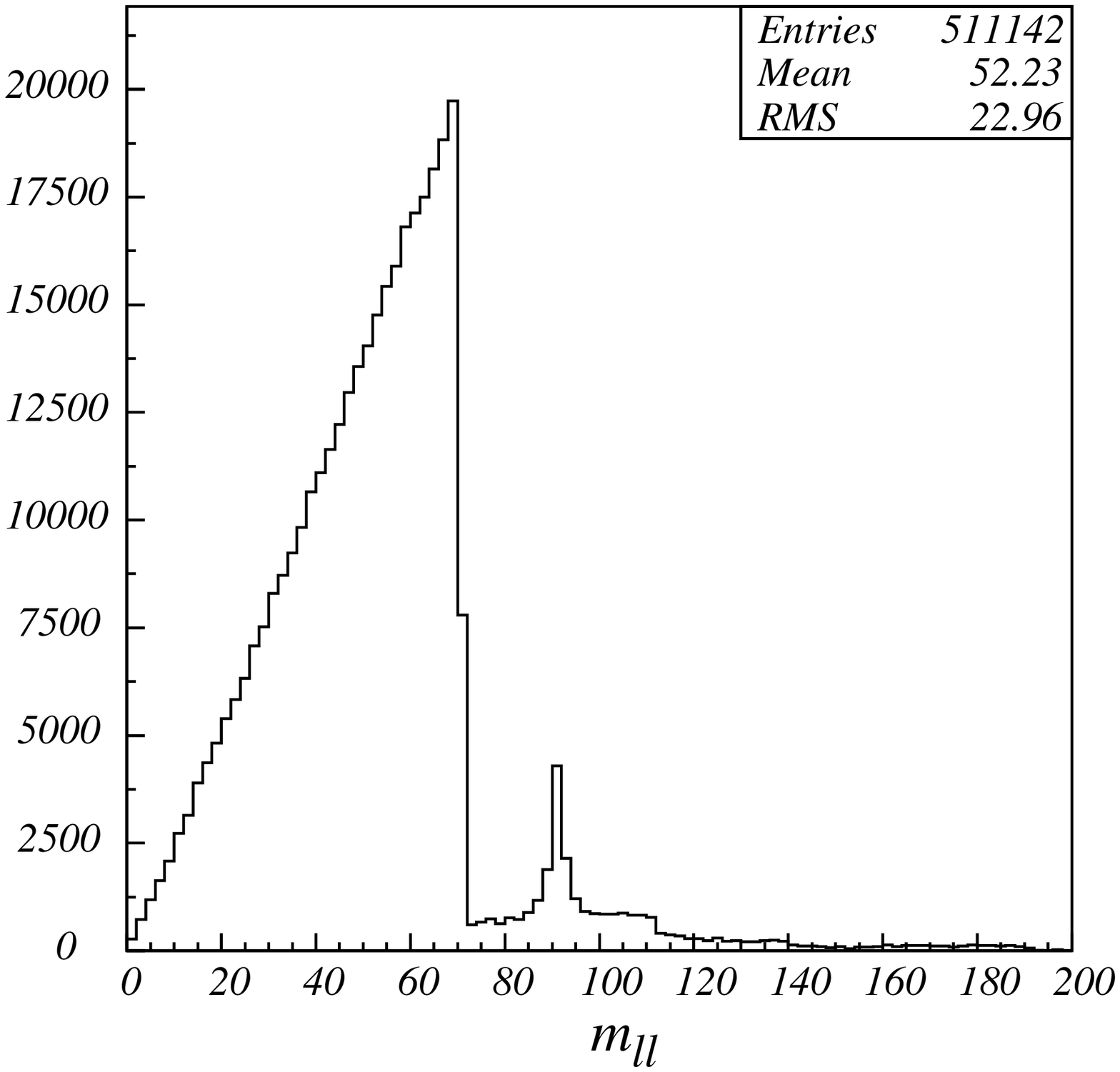}
\includegraphics[scale=0.3]{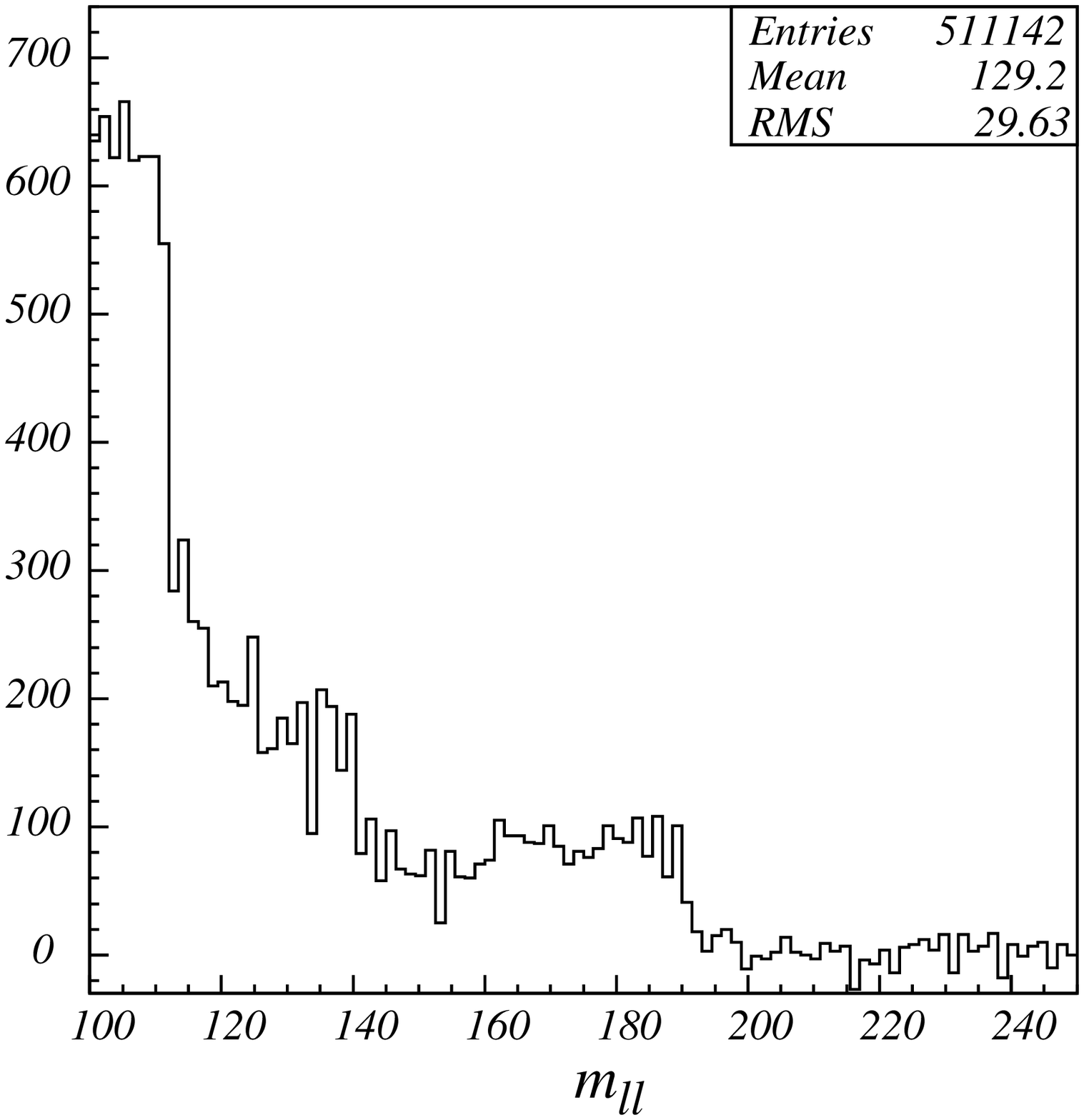}
\caption{Dilepton invariant mass distribution (in $\GEV$).}
\label{mll}
}

From the dilepton invariant mass distribution, edges of all decay modes except (3) in Table \ref{endp} are visible. The distribution of decay mode (3) is small because basically $\zii$ decays further through the golden mode (1) and results in four-lepton final states discussed below.

\subsection{Four-lepton channel}\label{4lsubsection}

Thus, we turn to consider four-lepton events. If these leptons really come from the cascade decay
\beq\
   \ziv\rar\sel\rar\zii\rar\ser\rar\lsp,\label{ziv_4l}
\eeq
they must form two OSSF pairs, and an invariant mass of one pair must be below the endpoint of mode (1) while another be lower than that of mode (3). This mode is very useful in the two senses. Basically, $\mwii\sim\mziv$, $\mwi\sim\mzii$ and $\msnu\sim\mel$ so that the endpoints from modes (2) and (3) are close. If endpoints of other modes are much different from these two, by looking at $m_{ll(3)}^{\rm max}$ from four-lepton events, one can easily pinpoint which edge from dilepton events is the edge of mode (2). Moreover, we can use $m_{ll(3)}^{\rm max}$ as a cross check for $\mziv$ and $\mel$ obtained from edges of mode (4) and (5).

We now consider the possibility to identify the decay mode (3) whose daughter $\zii$ subsequently decays via mode (1) producing four leptons in the final states. The following cuts are applied to select events:
\begin{itemize}
\item
exactly two pairs of OSSF leptons where each lepton has $p^l_T > 10 ~\GEV$ 
and $|\eta| < 2.5$,
\item
four leptons must be composed of exactly one $e^+e^-$ pair and 
one $\mu^+\mu^-$ pair.
\end{itemize}

\FIGURE[!ht]{
\hfill
\includegraphics[scale=0.3]{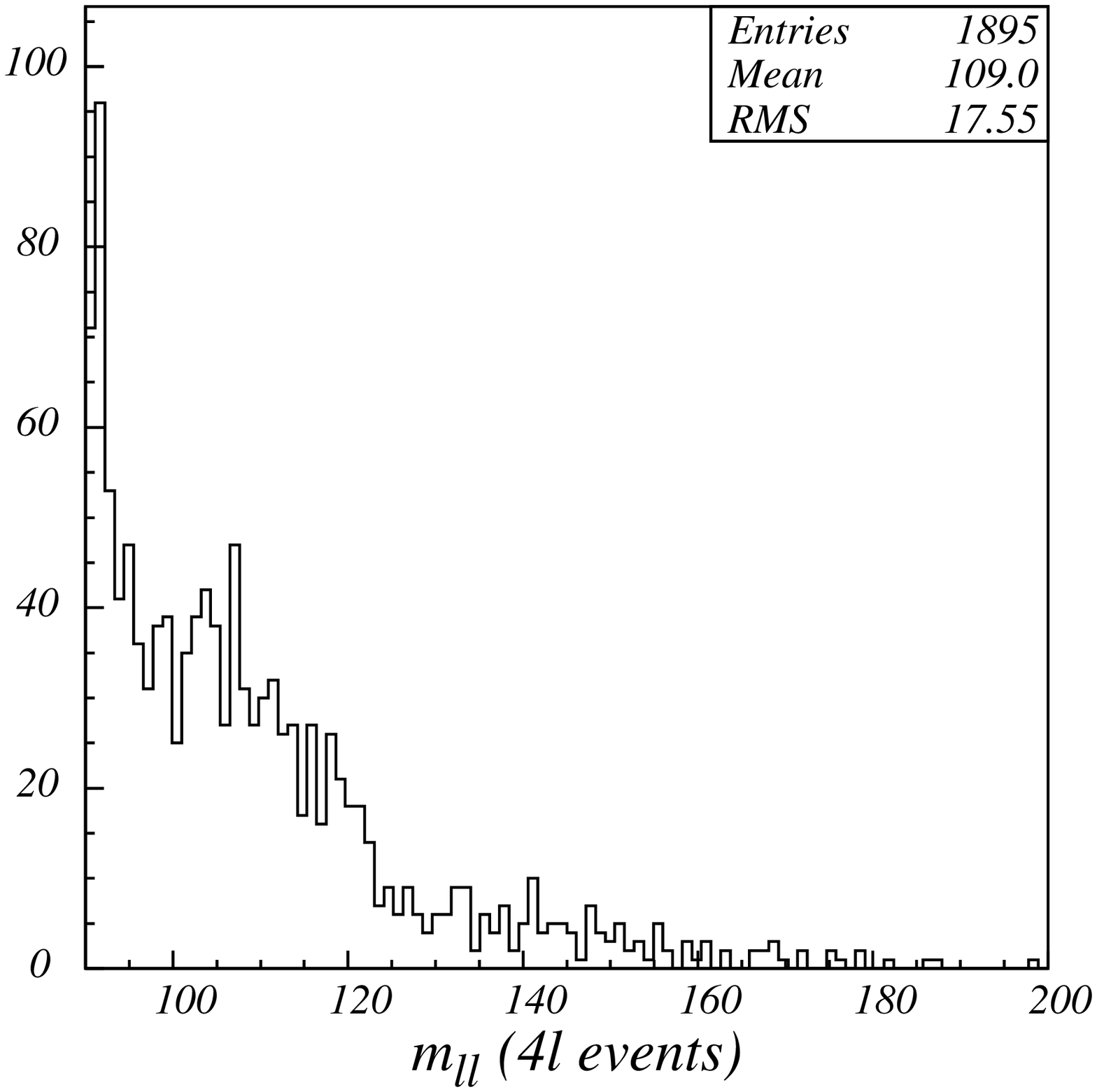}
\caption{Dilepton invariant mass $m_{ll}$ distribution (in $\GEV$) from four-lepton events with the requirement on the other lepton pair $m_{l'l'} < 70.7 ~\GEV$.}
\label{mll_4l}
}

The $m_{ee}$ and $m_{\mu\mu}$ are then calculated for each events. 
 After requiring that an invariant mass of one OSSF lepton pair must less than $70.7 ~\GEV$, distribution of the other pair is shown in Figure \ref{mll_4l}. Now the edge around 122 $\GEV$ shows up confirming that events are truly from the decay in Eq.~(\ref{ziv_4l}). For a consistency check, we also plotted the trilepton and four-lepton invariant mass distributions from events which pass the above cuts; the plots are shown in Appendix A. The fitted values of various endpoints are obtained by using linear fit function 
\beq
	f(M)=\left\{\begin{array}{ll}
	AM+B,& ~~~0\leq M\leq M^{\rm max}_{ll}\\
	0,& ~~~M> M^{\rm max}_{ll}
	\end{array}\right.
\eeq
smeared with a Gaussian and are listed in Table \ref{endpfit}.

\TABLE[!ht]{
\begin{tabular}{|c|c|c|}
\hline
Distribution & Expected endpoint & 
Fitted endpoint 
\\ \hline
\hline
$m_{ll(1)}$& 70.78 & 70.67 $\pm$ 0.01
\\ \hline
$m_{ll(2)}$ & 111.61 & 111.70 $\pm$ 0.21 
\\ \hline
$m_{ll(3)}$ & 122.41 & 121.93 $\pm$ 0.43 
\\ \hline
$m_{ll(4)}$ & 140.62 & 140.64 $\pm$ 0.26 
\\ \hline
$m_{ll(5)}$ & 190.46 & 191.05 $\pm$ 0.28 
\\ \hline
\end{tabular}
\caption{\footnotesize Fitted endpoints in $\GEV$}\label{endpfit}
}
%

\subsection{Mass Parameter Determination}

Once kinematical endpoints have been measured experimentally, relevant sparticle masses may be extracted. It is expected that $\msql , \mzii , m_{\tilde l_R} ,$ and $\mlsp$ can be reconstructed from $m^{max}_{ll}, m^{max}_{jll}, m^{max}_{jl}$ and $m^{min}_{jl}$ \cite{Allanach_JHEP2000}. Furthermore, if higher endpoints are visible, one would then be able to identify the decay of heavier neutralino and then resolve masses of $\ziv$ and $\tilde l_L$.

For a given set of $m^{max}_{ll}, m^{max}_{jll}, m^{max}_{jl}$ and $m^{min}_{jl}$ measurements, in principle, there may be different corresponding sets of sparticle masses \cite{Gjelsten}. For the set of endpoints (1) in Table \ref{endp}, there are two possible solutions as listed in Table \ref{invert_sol}. The second solution can be discarded by measuring $m^{min}_{jll}$, which differs by around $20~\GEV$. Alternatively, solution 2 does not have solution for $\mel$ which satisfies $m_{ll (4)}^{max}$ and $m_{ll (5)}^{max}$ simultaneously. 

\TABLE[!ht]{
\begin{tabular}{|c|c|c|c|c|}
\hline
& $\mlsp$ & $\mslr$ & $\mzii$ & $\msql$ 
\\ \hline
\hline
Solution 1 & 
114.70 & 130.38 & 197.82 & 665.19
\\ \hline
Solution 2 & 
178.09 & 244.03 & 265.06 & 747.22
\\ \hline
\end{tabular}
\caption{\footnotesize Sparticle mass ambiguities in $\GEV$}\label{invert_sol}
}

In order to estimate error for each SUSY particle mass, we generated a set of random numbers corresponding to a set of masses $\{\mlsp, \mzii, \mziv, \msql, \mel\}$. We then calculated a set of measurable quantities $\{ m_{jll}^{min}, m_{jll}^{max}, m_{jl}^{min}, m_{jl}^{max}, m_{ll (3,4,5)}^{max}\}$ and their chi-squared which is defined as
\beq\label{deltachi2}
	\Delta\chi^2&=&\sum_{\textrm{all observables}}
		\frac{(\textrm{Nominal value} - \textrm{Measured value})^2_i}{\sigma^2_i}.
\eeq

In the above definition, $\sigma_i$ includes both systematical error and statistical error. 
We employed systematical errors as listed in Table \ref{syserr} \cite{LHC/LC}. In our analysis, however, we used only statistical errors for $m_{ll (3,4,5)}^{max}$ since they dominate over systematical errors (see  Table \ref{endpfit}). Moreover, we fitted endpoints for $m_{jll}^{min}, m_{jll}^{max}, m_{jl}^{min}, m_{jl}^{max}$ distributions and found that even if they have good statistics, their fitted values curiously differ from central values, especially $m_{jll}^{max}$. We then just used their systematical errors in calculating $\Delta\chi^2$.

In addition, the negligibly smallness of both systematical and statistical errors for $m_{ll (1)}^{max}$ implies that $\mer$ will be measured rather precisely. Therefore, in the analysis, we took $m_{ll (1)}^{max}$ as an input and obtained $\mer$ for given values of $\{\mlsp, \mzii \}$. Results of 1-$\sigma$ error estimation for sparticle mass differences are shown in Table \ref{invt}. 

\TABLE[!ht]{
\begin{tabular}{|c|c|}
\hline
Distribution & Systematical error ($\GEV$)
\\ \hline
\hline
$m_{ll}^{max}$ & 0.08
\\ \hline
$m_{jll}^{max}$ & 4.3
\\ \hline
$m_{jll}^{min}$ & 2.0
\\ \hline
$m_{jl}^{max}$ & 3.8
\\ \hline
$m_{jl}^{min}$ & 3.0
\\ \hline
\end{tabular}
\caption{\footnotesize Estimated systematical errors for different endpoints (in $\GEV$).}\label{syserr}
}
%

\TABLE[!ht]{
\begin{tabular}{|c|c|c|}
\hline
Sparticle Mass & Central value & Estimated error 
\\ \hline
\hline
$\mlsp$ & 114.70 & ${}^{+6.7}_{-6.3}$
\\ \hline
$\mer -\mlsp$ & 15.68 & ${}^{+0.45}_{-0.49}$
\\ \hline
$\mzii -\mlsp$ & 83.12 & ${}^{+0.75}_{-0.62}$
\\ \hline
$\mel -\mlsp$ & 124.92 & ${}^{+0.65}_{-0.65}$
\\ \hline
$\mziv - \mlsp$ & 208.53 & ${}^{+0.77}_{-0.64}$
\\ \hline
$\msql -\mlsp$ & 551.19 & ${}^{+4.64}_{-4.47}$
\\ \hline
\end{tabular}
\caption{\footnotesize Central values and 1-$\sigma$ error estimation of relevant sparticle masses in $\GEV$.}\label{invt}
}
%

\section{Flipping Solutions}

Besides the one shown in Table \ref{invert_sol}, there is another kind of ambiguities which is related to parameter point identification especially when the ordering of $\mu$, $M_1$ and $M_2$ is shuffled. We now introduce flipping solutions. Flipping solutions are solutions among which masses of relevant sparticles (masses of left-handed squark, left- and right-handed sleptons, and three neutralinos which have significant gaugino component) are the same but ordering of  $M_1, M_2$ and $\mu$ parameters are different. The mass degeneracy among solutions results in the same endpoint positions from those sparticle decay. Therefore, only endpoint measurement is not enough to distinguish these solutions.
By reminding that $M_1 < M_2 < \mu$ for point A, we illustrate another two flipping solutions: $M_1 < \mu < M_2$ (point A2) and $\mu < M_1 < M_2$ (point A3); their relevant parameters are listed in Table \ref{solution23}. 

We fixed three masses of neutralinos which have significant gaugino component equal, as they are frequently produced from squark decays and would be measured rather precisely. For points A and A2, such three neutralino states are $\lsp, \zii$ and $\ziv$, so almost all masses except $\mziii$  for both points are degenerate ($\mziii |_{\rm point A}=279~\GEV$ and $\mziii |_{\rm point A2}=234~\GEV$). For point A3, however,
they are  $\lsp, \ziii$ and $\ziv$ instead. Therefore, we fixed $\mziii = \mzii |_{\rm point A}$ and $\mzii = 155~\GEV$. On the other hand, $\ziii$ for points A and A2 and $\zii$ for point A3 are nearly pure Higgsino, and they are not produced from squark decays, so that the experimental constraint would be weaker.

\TABLE[!ht]{
\begin{tabular}{|c|c|c|c||c|c|c|c|}\hline
& A & A2 & A3 && A & A2 & A3\\\hline\hline
$\mu$ & 271.33 & 226.06 & 146.21& & & &\\ \hline
$M_1$ & 122.49 & 125.66  & 187.46 &  $M_2$ & 230.89 & 272.86  & 291.47\\ \hline
\hline
$\mupl$ & 665.19 & 665.19 & 665.19 & $\mupr$ & 648.85 & 649.84 & 657.09\\ \hline
$\mdnl$ & 670.29 & 670.29 & 670.29 &$\mdnr$ & 642.47 & 641.67 & 643.68\\ \hline
$\mel$ & 239.62 & 239.62 & 239.62 & $\mer$ & 130.38 & 130.38 & 130.38 \\ \hline
$\mtauh$ & 238.89 & 270.32 & 292.60 & $\mtaul$ & 128.07 & 123.82 & 173.80 \\ \hline
$\msnu$ &  224.37 &  224.03 &  223.92 & $\mglu$ & 719.67 & 719.35 & 720.81\\ \hline
$\mwi$ & 196.30 & 193.34 & 133.27 & $\mwii$ & 321.62 & 320.88 & 319.79\\ \hline
$\mlsp$ & 114.70 & 114.70  & 114.70 & $\mzii$ & 197.82 & 197.82 & 155.17\\ \hline
$\mziii$ & 278.87 & 234.28 & 197.82 &$\mziv$ & 323.23 & 323.23 & 323.23 \\ \hline
\end{tabular}\\

%
%
\caption{\footnotesize Relevant parameters and sparticle masses in GeV for points A, A2 and A3 respectively.} \label{solution23}
}

In the followings, we discuss and compare some phenomenological signatures of all flipping solutions. We will show that different properties of  neutralinos essentially lead to discrepancy in both collider and low-energy LFV observables.

 \FIGURE[!ht]{
\includegraphics[scale=0.25]{ll_s1.eps}
\includegraphics[scale=0.25]{ll_s1_fig2.eps}\\
\includegraphics[scale=0.25]{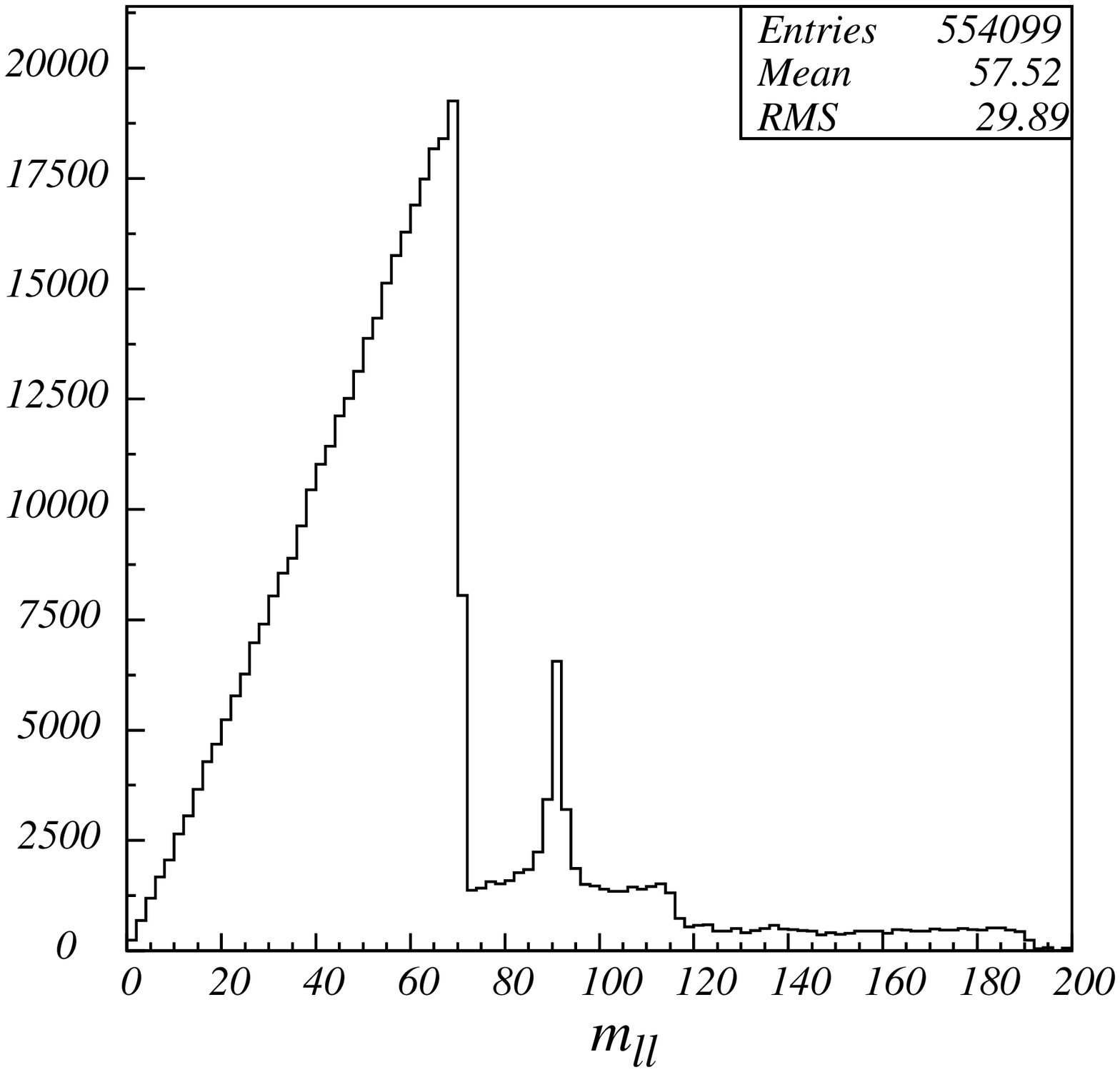}
\includegraphics[scale=0.25]{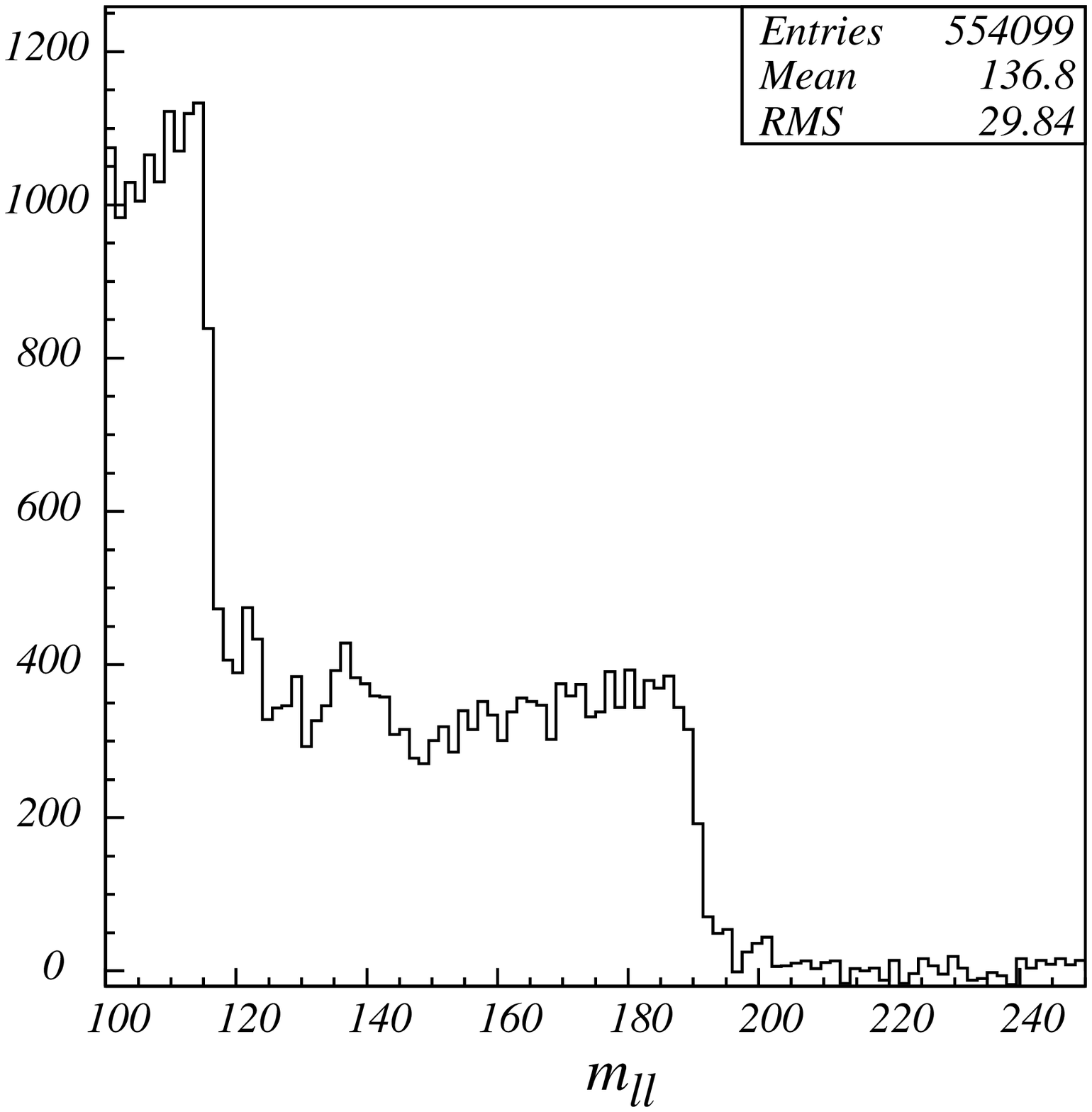}\\
\includegraphics[scale=0.25]{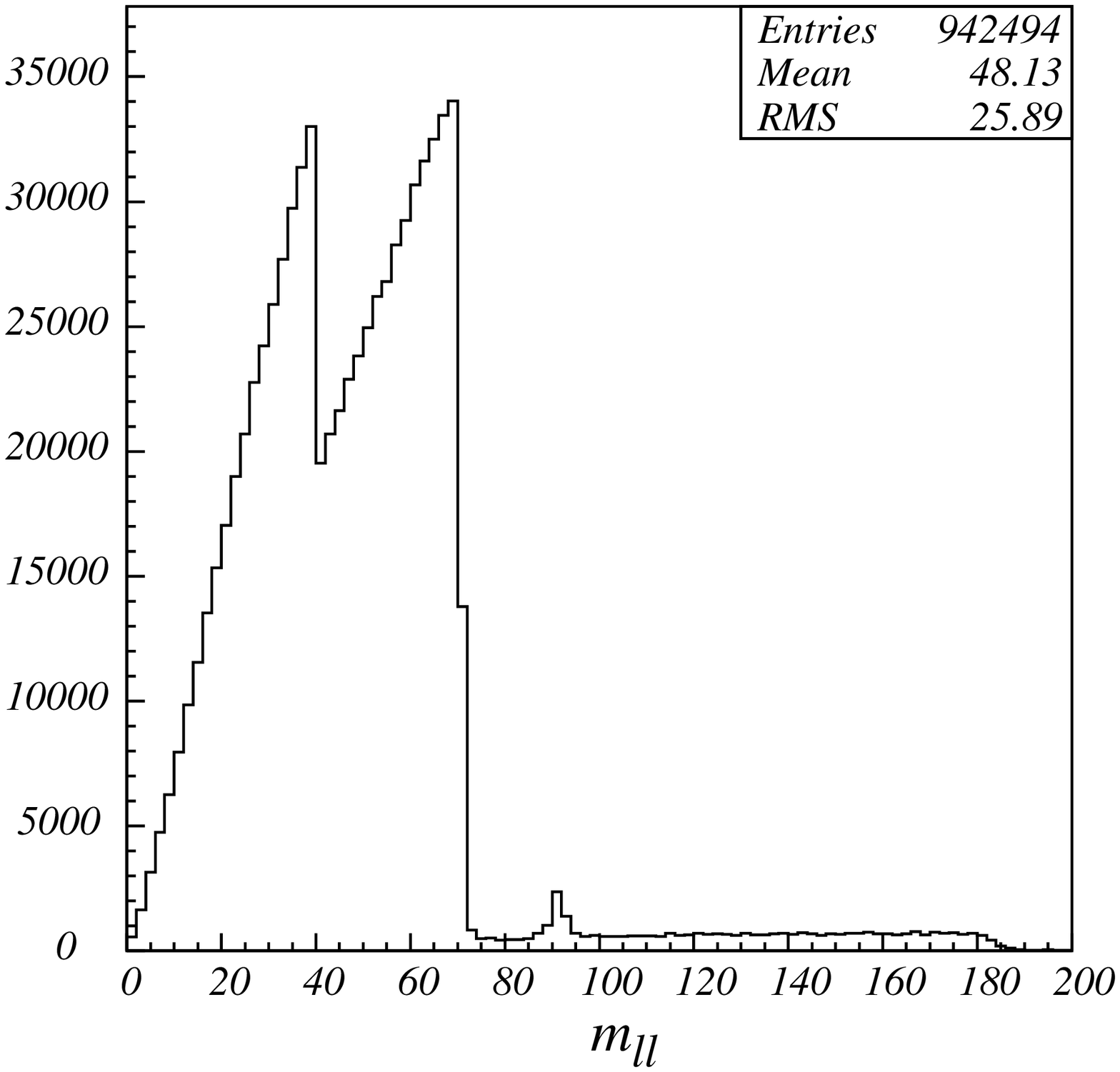}
\includegraphics[scale=0.25]{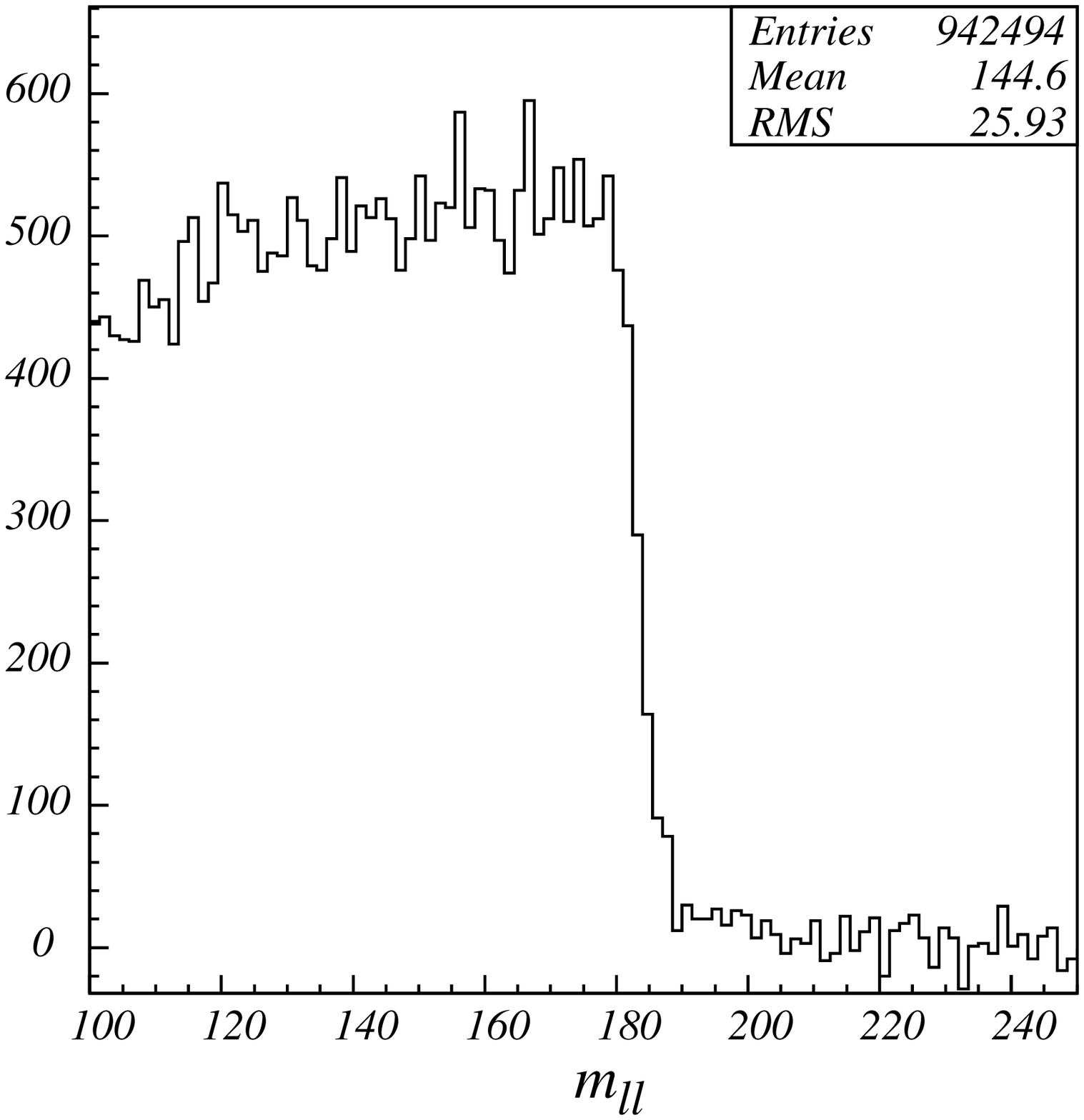}\\
\vspace{-1cm}
\caption{OSSF dilepton invariant mass distribution for points A, A2 and A3 respectively.  Right panels show a close-up of higher endpoints.}
\label{ll_s123}
}
\subsection{Branching Ratios}

First of all, we show the OSSF dilepton distributions in Figure \ref{ll_s123}. The total number of generated events is $5\times 10^6$ for all points. This corresponds to about 300 $\fb^{-1}$ for points A and A2 and 250 $\fb^{-1}$ for point A3. The difference between distribution for points A and A2 is conspicuous. Since $\wiipm$ and $\ziv$ are Wino-like at point A2, $Br(\sql\rar\wii /\ziv)$ and $Br(\wii /\ziv\rar\sll/\snu)$ are enhanced (see Table \ref{br_s1s2}). These result in the $Z$ peak and the height of edges around $110~\GEV$ (from $\wii\rar\snu\rar\wi$) and $190~\GEV$ (from $\ziv\rar\sel\rar\lsp$) for A2 at least twice the size of that for point A. 

\TABLE[!ht]{
\begin{tabular}{|c|c|c||c|c|c||c|c|c|}
\hline
& A & A2 & &A & A2 & &A & A2\\ \hline
$\sul\rar\zii$ & 25.4  &  15.4 
&$\sur\rar\zii$ & 4.1 & 9.6
&$\zii\rar\ser$ & 13.5 & 14.1\\ \hline

$\sul\rar\ziv$ & 7.8  &  17.8 
&$\sdr\rar\zii$ & 4.1 & 9.6
&$\wii\rar\snu$ & 7.4 & 12.9 \\ \hline  

$\sul\rar\wii$ &13.1  &  33.2 
&$\ziv\rar Z+\zii$ & 0.2 & 0.2 
&$\wii\rar Z+\wi$ & 16.1 & 13.4\\ \hline

$\sdl\rar\zii$  & 21.8 & 10.4 
&$\ziv\rar Z+\lsp$ & 1.4 & 0.3
&$\snu\rar\wi$ & 26.1& 20.4\\ \hline

$\sdl\rar\ziv$ & 9.7   &  20.5 
&$\ziv\rar\sel$ & 2.3 & 4.2 
&$\sel\rar\zii$ & 26.5 & 25.6\\ \hline

$\sdl\rar\wiic$ & 23.1& 45.2  
&$\ziv\rar\ser$ & 0.7 & 0.3 
&$\sel\rar\lsp$ & 35.9 & 47.8\\ \hline

\end{tabular}
\caption{\footnotesize Comparison between relevant branching ratios in \% for points A and A2.}\label{br_s1s2}
}

The enhancement of the branching ratios $Br(\sql\rar\wii)$ and $Br(\wii\rar\snu)$ can be  understood analytically. Consider the $\widetilde\chi_2^{\pm}$-lepton-slepton interaction
\beq
	{\mathcal L}&\propto&
		g\sin\phi_R(\tilde\nu^*\overline{\widetilde\chi}^-_2P_L e 
		+\bar eP_R\widetilde\chi^-_2\tilde\nu )
	 	+g\sin\phi_L(\sel^*\overline{\widetilde\chi}^+_2P_L \nu
		+\bar\nu P_R\widetilde\chi^+_2\sel )	.		
\eeq
Here $P_{L,R}$ stands for left-, right-handed projection operator and the mixing angles $\phi_L$ and $\phi_R$ are given (with two-fold ambiguity) by
\beq
	\tan 2\phi_L &=& \frac{2\sqrt{2}m_W(M_2\cos\beta+\mu\sin\beta)}
		{M_2^2-\mu^2-2m_W^2\cos 2\beta}\,,\\
	\tan 2\phi_R &=& \frac{2\sqrt{2}m_W(M_2\sin\beta+\mu\cos\beta)}
		{M_2^2-\mu^2+2m_W^2\cos 2\beta}.
\eeq
The interaction $\widetilde\chi_2-\tilde\nu-e$ and, analogously, $\widetilde\chi_2-\tilde u-d$ are proportional to $\sin\phi_R$ which becomes larger when $M_2$ and $\mu$ are inverted so that $\mu < M_2$, for moderate $\tan\beta$.

Point A3 can also be distinguished from the others.
The important key is that, for point A3, mass difference $\mwii-\mwi$ is close to $\mziv-\mlsp$, in addition to $\mziv-\mzii$. Hence the dislocation of $m^{max}_{ll}(\wii\rar\snu\rar\wi)$ will serve as an indicator of flipping solution. In Figure \ref{ll_s123}, the highest endpoint around $180~\GEV$ for point A3 actually comes from the decay $\wii\rar\snu\rar\wi$, while the distribution for $\ziv\rar\sel\rar\lsp$ ($m^{max}_{ll}\sim 190~\GEV$) is much smaller. 

In addition, the study of four-lepton events in analogous to subsection \ref{4lsubsection} reveals the endpoint from the decay chain $\ziv\rar\sel\rar\ziii$ $(m^{max}_{ll}\sim 120~\GEV$) where $\ziii$ subsequently undergoes the cascade decay $\ziii\rar\ser\rar\lsp$. By utilizing all endpoint information from four-lepton events, $\ziv$ and $\sel$ masses can be extracted and $m^{max}_{ll}(\ziv\rar\sel\rar\lsp)\sim 190~\GEV$ can be calculated. To this end, one can confirm that the $m_{ll}$ endpoint at $180~\GEV$ does not come from $\ziv$ decay.

A very striking feature of its double peak distribution below the $Z$ peak in Figure \ref{ll_s123} seems to make $\mu < M_1 < M_2$ case very easily distinguishable
 from other cases. 
 However, this actually depends on our model parameter choice, i.e. $\mer-\mlsp\ll\mzii-\mer$. The distribution receives contribution to the outer edge (around $70~\GEV$) from the Bino-like $\ziii$ cascade decay $\ziii\rar\slr\rar\lsp$ and to the inner edge (around $40~\GEV$) from the Higgsino-Bino mixed $\zii$ cascade decay $\zii\rar\slr\rar\lsp$. 
 Consider the case that our true model point is the solution 2 in Table \ref{invert_sol}. Its $\mu < M_1 < M_2$ flipping solution (point A4) has $\mzii < \mer$ and an unusual double peak distribution disappears. We show mass spectrum of the point in Table \ref{solution4} and $m_{ll}$ distribution in Figure \ref{mll_a4}.

\TABLE[!ht]{
Point A4:
\begin{tabular}{|c|c||c|c||c|c||c|c|}\hline
$\mu$& 205.64 & $M_1$ & 253.75  &  $M_2$& 414.91 & &\\ \hline
\hline
$\mupl$ & 747.54 & $\mupr$ & 696.83 &
$\mdnl$ & 752.02  &$\mdnr$ & 680.72 \\ \hline
$\mel$ & 389.37  & $\mer$ & 243.92 &
$\mtauh$ & 387.60 & $\mtaul$ & 233.51  \\ \hline
$\msnu$ &  380.13 & $\mglu$ & 766.77 &
$\mwi$ & 198.64 & $\mwii$ & 435.96 \\ \hline
$\mlsp$ & 179.04 & $\mzii$ & 212.99 &
$\mziii$ & 265.21 &$\mziv$ & 440.05  \\ \hline
\end{tabular}
\hspace{1.5cm}
\vspace{.5cm} 
\caption{\footnotesize Relevant parameters and sparticle masses in GeV for point A4.} \label{solution4}
}
%

\FIGURE[!ht]{
\includegraphics[scale=0.3]{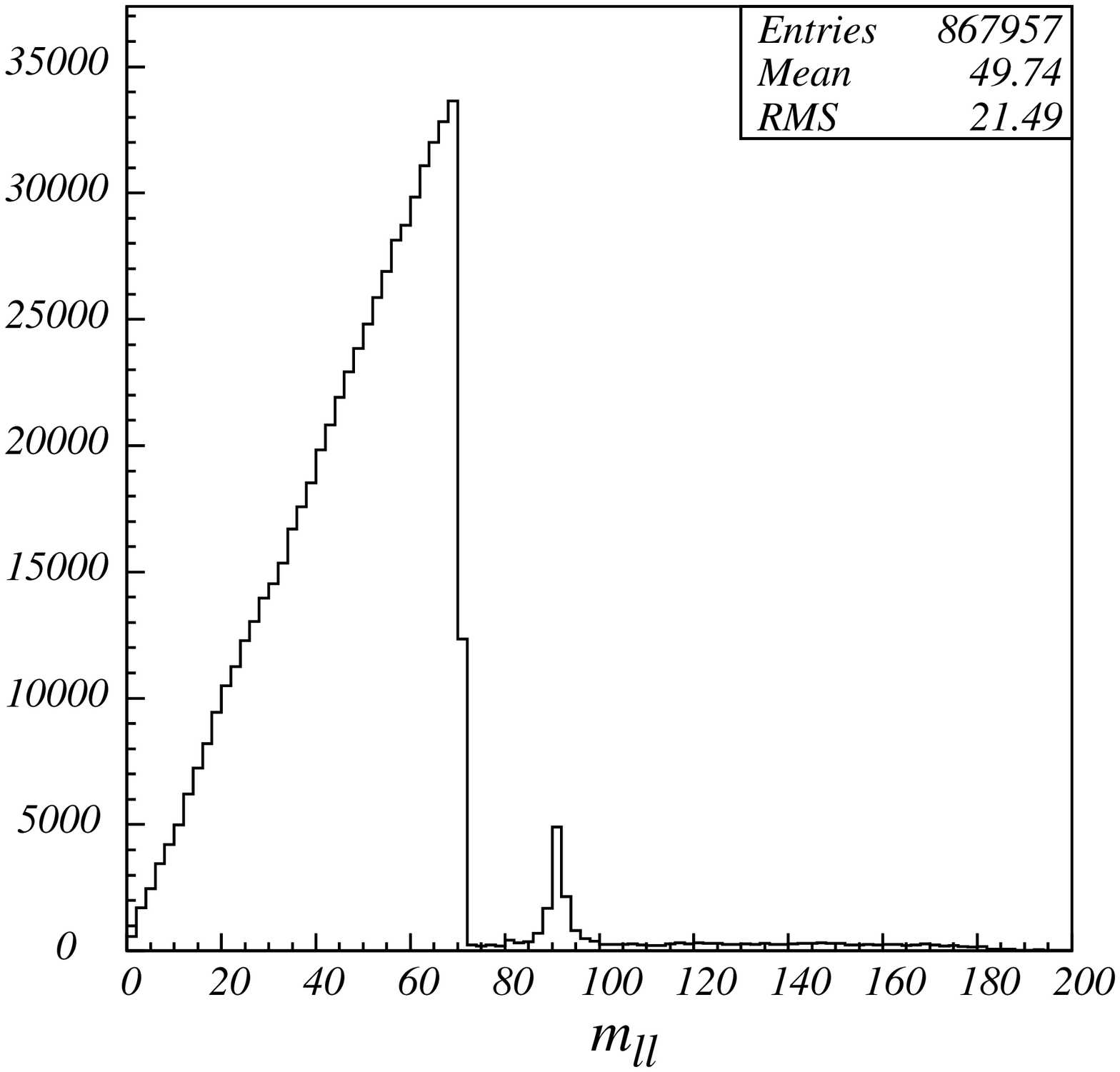}
\caption{OSSF dilepton invariant mass distribution for point A4.}\label{mll_a4}
}
%

\subsection{2 jets + $\esla_T$ Signature}

The distinctive Higgsino-like property of $\lsp$ for very small $\mu$ solution makes it possible to distinguish one from intermediate-to-large $\mu$ solutions. More specifically, it is clearly presented in the differences in the decay branching ratio
\beq
	Br(\sqr\rar\lsp)=\left\{\begin{array}{l}
	0.94\hspace{2cm}\textrm{point A}\\
	0.89\hspace{2cm}\textrm{point A2}\\
        0.16\hspace{2cm}\textrm{point A3}\end{array}~~~.\right.\non
\eeq

When both pair-produced right-handed squarks decay directly into $\lsp$, their signature is two high $p_T$ jets $+$ large $\esla_T$ and no isolated lepton in the final states. Here, we first count number of jets. We select events by employing the following cuts:\\
\indent 
(c1) more than 2 jets with $|\eta_i| < 2.5$ and transverse momentum $p_{T,1}> 250~\GEV, p_{T,2}> 200~\GEV$,\\
\indent
(c2) no isolated lepton with $p_T$ greater than $10~\GEV$,\\
\indent
(c3) no tagged b-jet,\\
\indent
(c4) $\esla_T> 250~\GEV$. \\

The number distribution of jet with $p_T> 50~\GEV$, $n_{j50}$, after cuts is shown in  Figure \ref{njets}. Note that the distribution for point A2 is very similar to that for point A and therefore not shown here. There are obviously two differences between these two plots. Firstly, high jet-$p_T$ cut (c1) and lepton cut (c2) reduce the number of events for point A3 so substantially that it is an order of magnitude smaller. Secondly, point A3 has smaller fraction of events for $n_{j50}=2$ and larger fraction when $n_{j50} > 4$. These all indicate that squark tends to decay via a longer cascade decay which produces leptons or several softer jets in the final states. It further implies that squarks would have considerable decay branching ratio into heavier inos and the LSP has small gaugino components, i.e. Higgsino-like. To this end, one can infer that point A3 has small $\mu < M_1, M_2$ . 

In the same plots, the dashed line represents distribution of $pp\rar\sqr\sqr\rar\lsp\lsp$ events. Even though there must have only two jets at parton level, the initial state radiation can contribute to the high $p_T$ third and forth jets. The difference between the solid and dashed line receives contributions mostly from left-handed squark or gluino (associated) production.

\FIGURE[!ht]{
\includegraphics[scale=0.25]{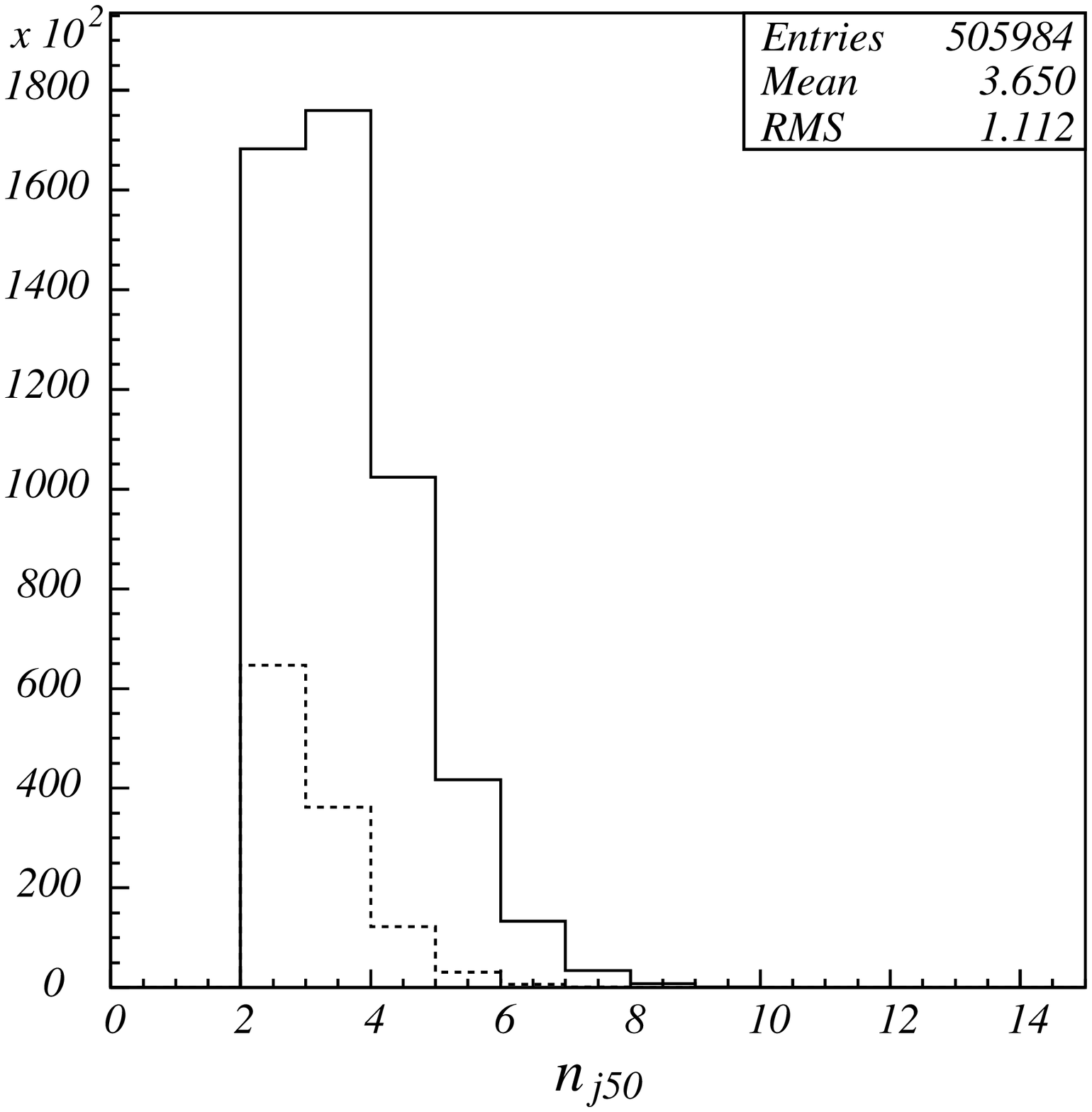}
\includegraphics[scale=0.25]{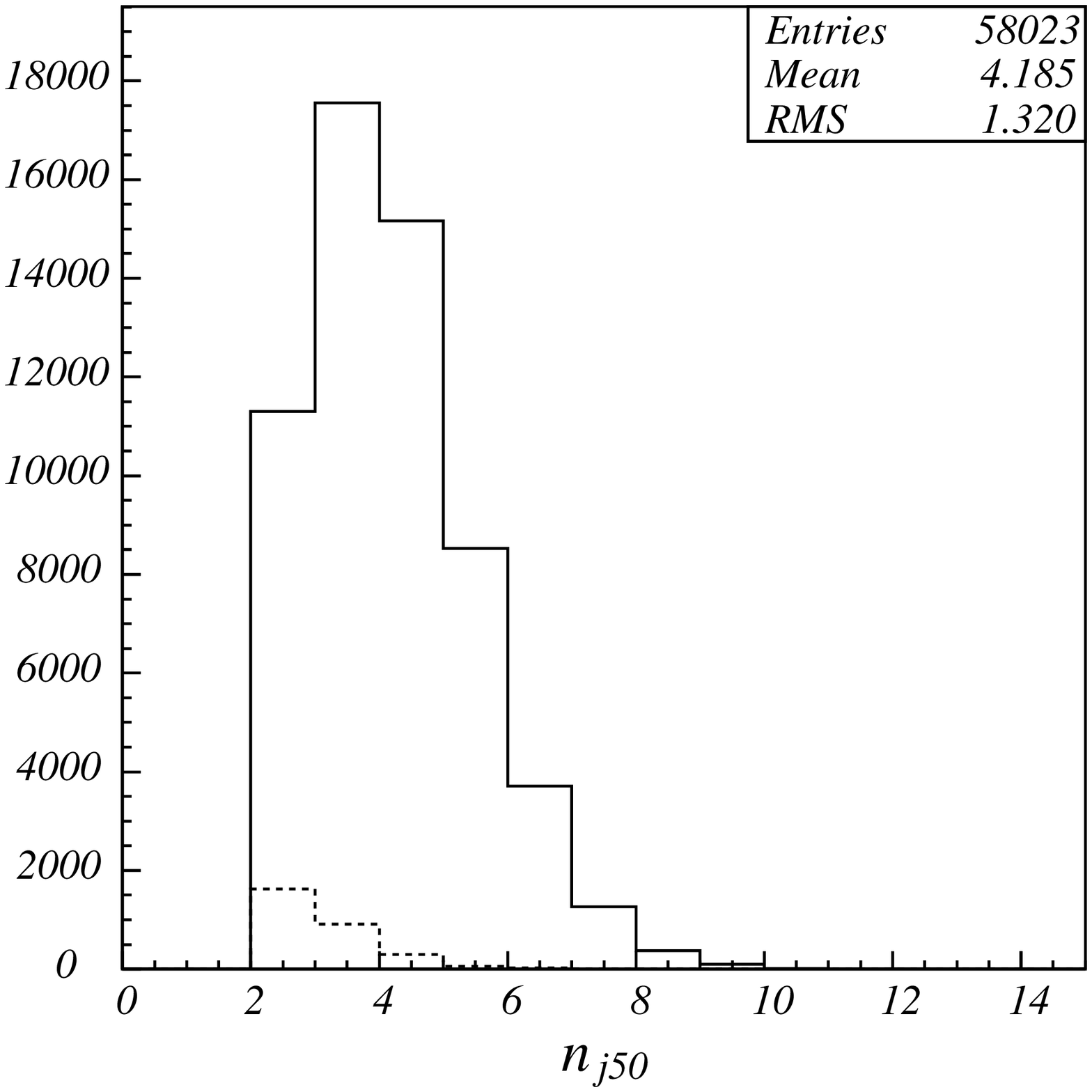}
\caption{The number distributions of jet with $p_T> 50 \GEV$ after the event cuts for point A and A3, respectively. The dashed line represents the distribution for $pp\rar\sqr\sqr\rar\lsp\lsp$. Notice that the vertical axis of the left-handed figure is multiplied by 100.}
\label{njets}
}
%

\FIGURE[!ht]{
\includegraphics[scale=0.25]{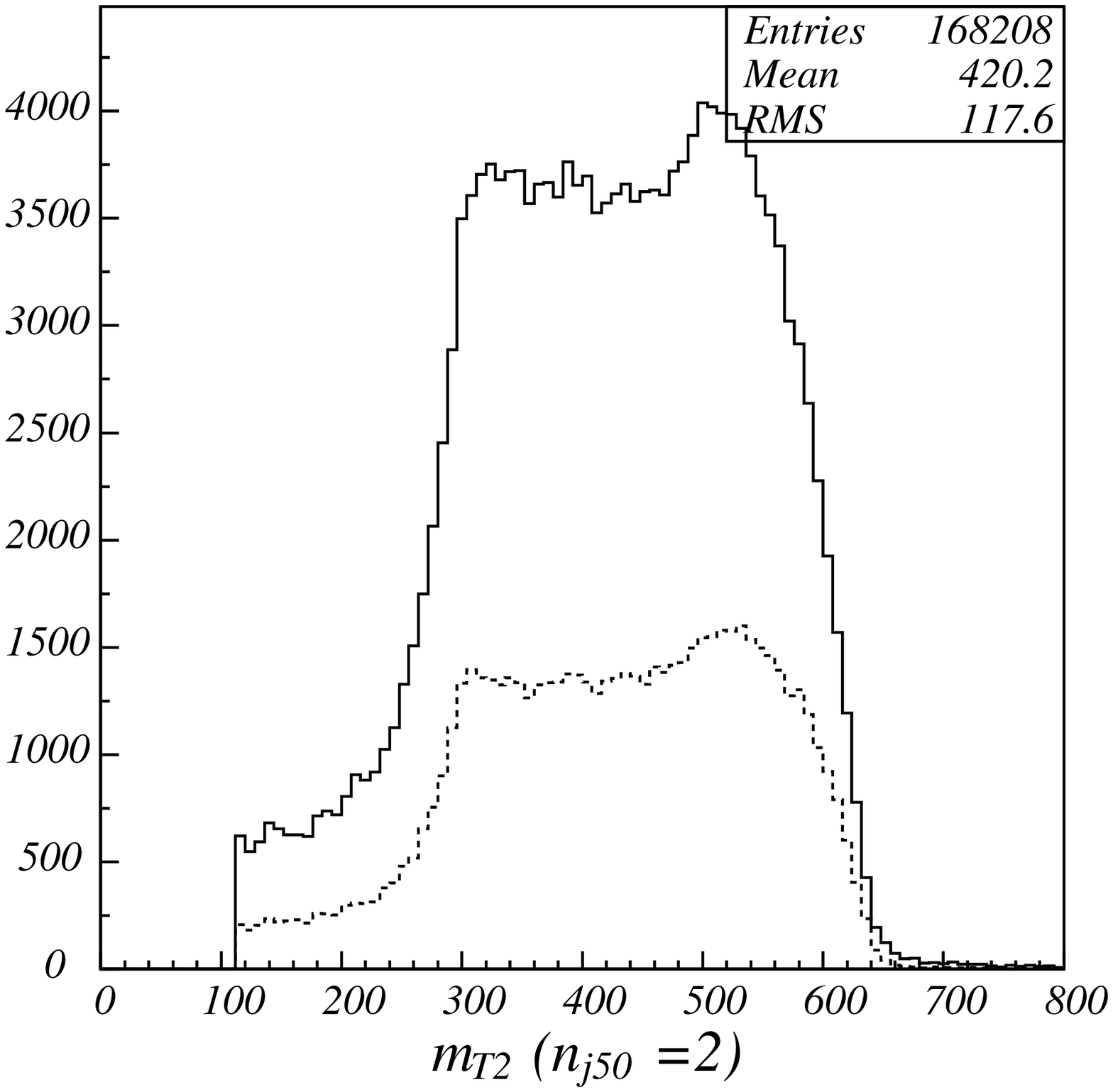}
\includegraphics[scale=0.25]{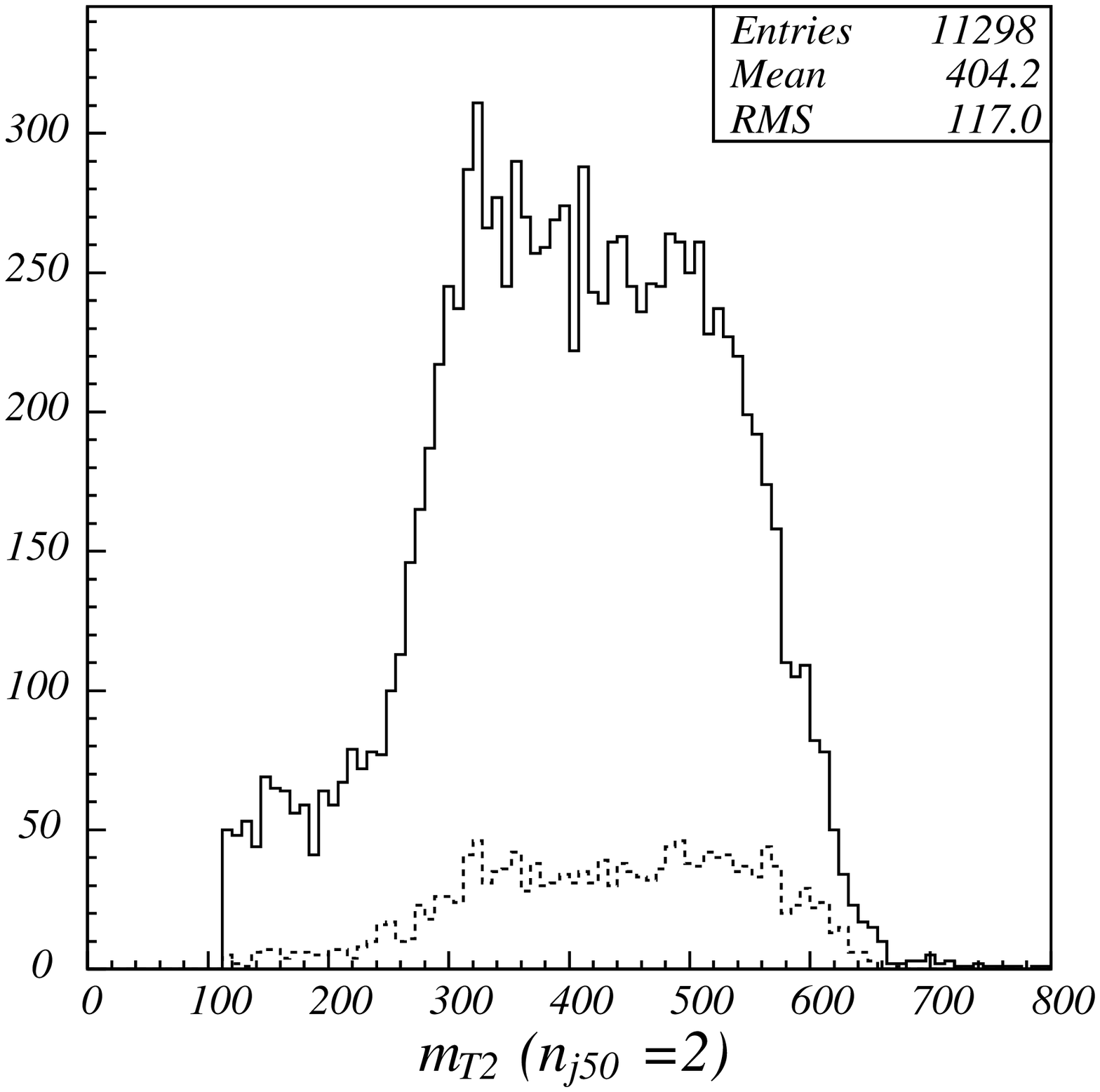}
\caption{$m_{T2}$ distribution for $n_{j50}=2$. The left and right panels are for point A and A3 respectively. The dashed line represents the distribution for $pp\rar\sqr\sqr\rar\lsp\lsp$.}
\label{mt2}
}

The method of counting jet numbers, however, is not so reliable. We need more concrete evidence to confirm the conclusions about small $Br(\sq\rar\lsp)$ and hence small value of $\mu$. This can be achieved by the so-called $m_{T2}$ method \cite{mt_two}. The $m_{T2}$ variable is defined in event-by-event basis as

\beq
	m_{T2}(\mathbf{p}_T^{j_1},\mathbf{p}_T^{j_2},\mathbf{p} {\!\!\!/}_T^{\rm miss};m_{\rm test}) 
\equiv 
\min_{\mathbf{p} {\!\!\!/}_T^{\alpha} + \mathbf{p} {\!\!\!/}_T^{\beta} = \mathbf{p} {\!\!\!/}_T^{\rm miss}}
\left[
\max \left\{
m_T(\mathbf{p}_T^{j_1},\mathbf{p} {\!\!\!/}_T^{\alpha};m_{\rm test}),
m_T(\mathbf{p}_T^{j_2},\mathbf{p} {\!\!\!/}_T^{\beta};m_{\rm test}) 
\right\}
\right]
\eeq
where $m_{\rm test}$ is a test mass 
and the minimization is performed over all possible splitting $\mathbf{p} {\!\!\!/}_T^{\alpha} + \mathbf{p} {\!\!\!/}_T^{\beta} = \mathbf{p} {\!\!\!/}_T^{\rm miss}$. The transverse mass $m_T$ is defined as 
\beq
m_T^2(\mathbf{p}_T^j,\mathbf{p} {\!\!\!/}_T^{\alpha};m_{\rm test})
\equiv m_j^2 + m_{\rm test}^2
+ 2\left( E_T^j E {\!\!\!/}_T^\alpha -  \mathbf{p}_T ^{j}\cdot\mathbf{p} {\!\!\!/}_T^{\alpha}\right).
\eeq

The $m_{T2}$ variable has the property 
\beq
	m_{T2}(m_{\rm test}=\mlsp)\leq\msqr,
\eeq
for $pp\rar\sqr\sqr\rar\lsp\lsp$ events, so that one can determine $\msqr$ from the endpoint measurement of $m_{T2}$ distribution.

Figure \ref{mt2} shows $m_{T2}$ distributions for $n_{j50}=2$. Here, $m_{\rm test}$ is taken to be the nominal value of $\mlsp$. Clearly, both $m_{T2}$ distributions with $n_{j50}=2$ have endpoint about the correct $\sqr$ masses. However, the distribution for point A has sharp edge and events near the edge come mostly from the true cascade while that for A3 is contaminated mostly by contribution from $\sql$ production.

In order to estimate a statistical significance of signal over SM background, we adopted a  set of event selection cuts from \cite{atlas_ep}: 
\begin{itemize}
\item
$\esla_T> {\rm max}(200~\GEV, 0.25M_{eff})$ and $M_{eff}>500~\GEV$,
\item
two jets with $p_T> {\rm max}(200~\GEV, 0.25M_{eff}), |\eta|<1$ and $\Delta R>1$,
\item
no additional jet with $p_T> {\rm min}(200~\GEV, 0.15M_{eff})$,
\item 
no isolated leptons and no tagged b-jets,
\item
transverse sphericity $S_T>0.2$.
\end{itemize}

It should be noted here that the above two hardest $p_T$ jet cut, $p_T> {\rm max}(200~\GEV, 0.25M_{eff})$, distorts shape of $m_{T2}$ distribution. This is because the event near $m_{T2}$ endpoint corresponds to the configuration where two jets go in the same direction with $p_{T,j}\sim 0.25M_{eff}$. Therefore, this cut kills significant numbers of events near the endpoint.

In Figure \ref{mt2_ptcut}, we just show how $m_{T2}$ distribution is distorted. 
On the left panel, $m_{T2}$ distribution for point A is shown when the above cuts are applied at $\sim 30~fb^{-1}$ of integrated luminosity. It is an $M_{eff}$ dependence of the cut that selects events with rather high $p_T$ jets. If this cut is relaxed to be $p_{T,1}> 250~\GEV, p_{T,2}> 200~\GEV$, in the right panel, the peak of $m_{T2}$ distribution leans more toward its edge as it is supposed to be. Moreover, the number of events passing the cuts are about three times larger with increasing fraction of $pp\rar\sqr\sqr\rar\lsp\lsp$ events near endpoint.

\FIGURE[!ht]{
\includegraphics[scale=0.25]{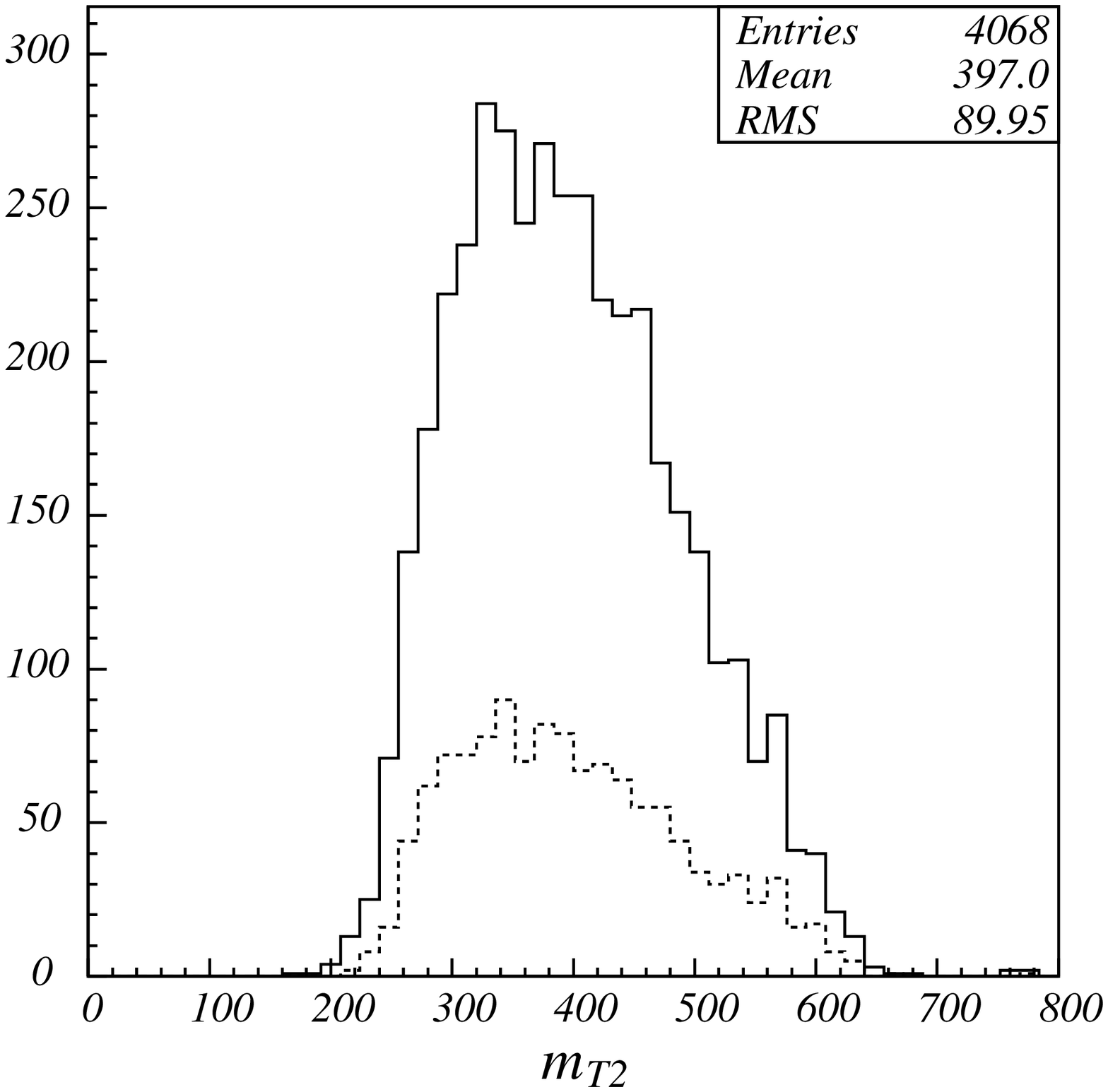}
\includegraphics[scale=0.25]{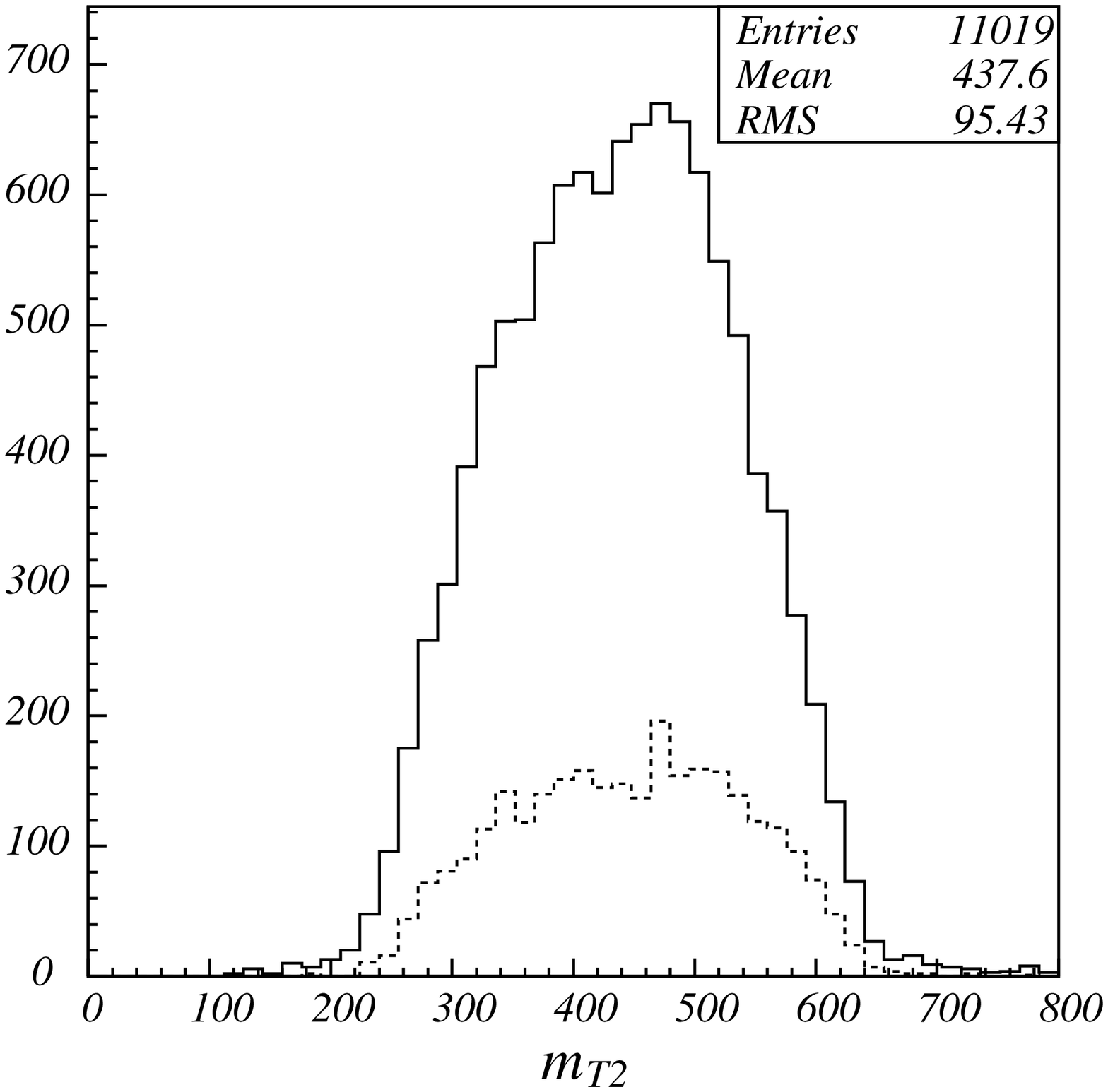}
\caption{$m_{T2}$ distribution of point A at $30~fb^{-1}$ of integrated luminosity. In the left panel, selection cuts are taken from \cite{atlas_ep}. The two hardest $p_T$ jet cut is different for the right panel as described in the text. Again, the dashed line represents the distribution for $pp\rar\sqr\sqr\rar\lsp\lsp$.}
\label{mt2_ptcut}
}

The signal-to-background ratio and signal statistical significance of events passing above cuts for $10~fb^{-1}$ of integrated luminosity are given in Table \ref{mt2_s2b_ratio} for points A and A3. The background statistics is taken from  \cite{atlas_ep}, page 1630.
The signal statistic of point A3 is so poor that its $m_{T2}$ distribution is submerged in SM background. This makes it even easier to distinguish between two points.

\TABLE[!ht]{
\begin{tabular}{|c|c|c|c|c|}
\hline
~Point~&~No. of Signal~&~No. of SM Background~&
~S/${\rm B}_{SM}$~ & ~S/$\sqrt{{\rm B}_{SM}}$~
\\ \hline
\hline
A & 1341 & 180 & 7.5 & 100.0
\\\hline
A3 & 133 & 180 & 0.7 & 9.9
\\ \hline
\end{tabular}
\caption{\footnotesize The number of signal and SM background events passing the selection cuts described in the text, signal-to-SM background ratio, and signal statistical significance for points A and A3 at $10~fb^{-1}$ of integrated luminosity.}
\label{mt2_s2b_ratio}
}

Summarily, by investigating the $m_{T2}$ distribution and branching ratio of heavier ino decay, the region to which the $\mu$ parameter belongs can be ascertained. 

\subsection{Charge Asymmetry}

In this subsection, we will illustrate that a charge asymmetry between the $jl^+$ and $jl^-$ invariant mass distributions can help lift degeneracy between points A and A2.
This method was firstly proposed in \cite{Barr} where the decay $\sql\rar\tilde{\chi}^0_2\rar\tilde{l}_\beta$ is studied in the case that slepton is purely right-handed; later the left-right mixing was taken into account in \cite{Goto}. In our study, we assume that left-right slepton mixing is negligible. We now briefly explain the method following prescription in \cite{Barr}. 

Consider the cascade decay $\tilde q_\alpha\rar q\tilde{\chi}^0_i\rar q l_{near}\tilde l_\beta$ where $l_{near}$ denotes lepton from $\tilde\chi^0_i$ decay and $\alpha,\beta=L,R$. The $ql_{near}$ invariant mass is given by a simple kinematics expression
\beq
	(m_{ql_{near}})^2 &=&(m_{ql_{near}}^{max})^2s^2_{\theta/2}
\eeq
where $s_{\theta/2}\equiv\sin(\theta/2)$ and $\theta$ is the angle between quark and lepton momenta in $\tilde{\chi}^0_i$ rest frame. Due to the chirality structure of quark-squark-neutralino coupling, $\tilde\chi^0_i$ is polarized. Its polarization alters the angular distribution of its daughter lepton and hence the angular distribution of $ql_{near}$ invariant mass. Mathematically, the probability density function receives extra angular-dependent factors from the chirality projector. For the case $\alpha\ne\beta$, the probability density is given by

\beq\label{angular_dis}
	\frac{dP}{ds_{\theta/2}}=\left\{\begin{array}{l}
	4s^3_{\theta/2}\hspace{3cm}
		{\rm for~~} ql^+_{near}{~\rm or~}\bar ql^-_{near}\\
        4s_{\theta/2}(1-s^2_{\theta/2})\hspace{1.3cm}
        		{\rm for~~}ql^-_{near}{~\rm or~}\bar ql^+_{near}\end{array}~~~.\right.
\eeq
If $\alpha=\beta$, the density function for $ql^-_{near}$ and $\bar ql^+_{near}$ are inverted. Based on the fact that valence quarks have harder PDFs than sea quarks,  squarks would be produced via $\tilde{q}\tilde{g}$ production more than anti-squarks. In the following, we therefore assume that high-$p_T$ jet represents quark rather than anti-quark. 

Define charge asymmetry
\beq
	A\equiv \frac{s^+-s^-}{s^++s^-}\hspace{1cm}{\rm where}\hspace{.8cm}
		s^\pm=\int^{m^{max}_{ql^\pm}}_{(m^{max}_{ql^\pm})/2}d\sigma(m_{ql^\pm}).
\eeq
We obtain
\beq
	A=\left\{\begin{array}{l}
		+\hspace{2cm}{\rm for~~} \alpha\ne\beta\\
		-\hspace{2cm}{\rm for~~} \alpha=\beta\end{array}.\right.
\eeq

For the celebrated decay mode $\sql\rar\zii\rar\slr$, $l_{near}$ with positive charge rather than negative charge favors to go to the opposite direction to the quark jet and constitutes events near endpoint. The charge asymmetry for this case is then positive.

\FIGURE[!ht]{
\includegraphics[scale=0.25]{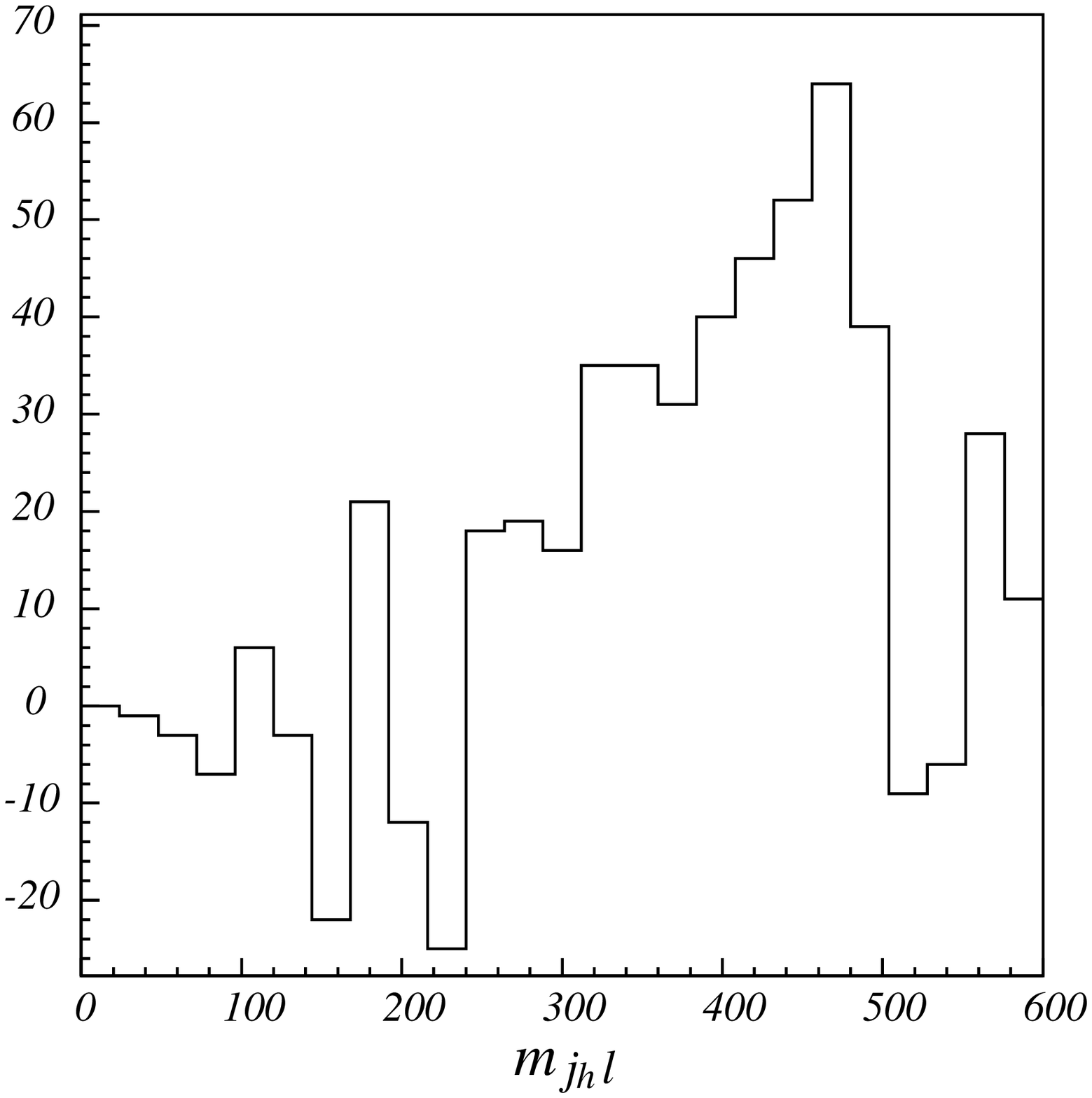}
\includegraphics[scale=0.25]{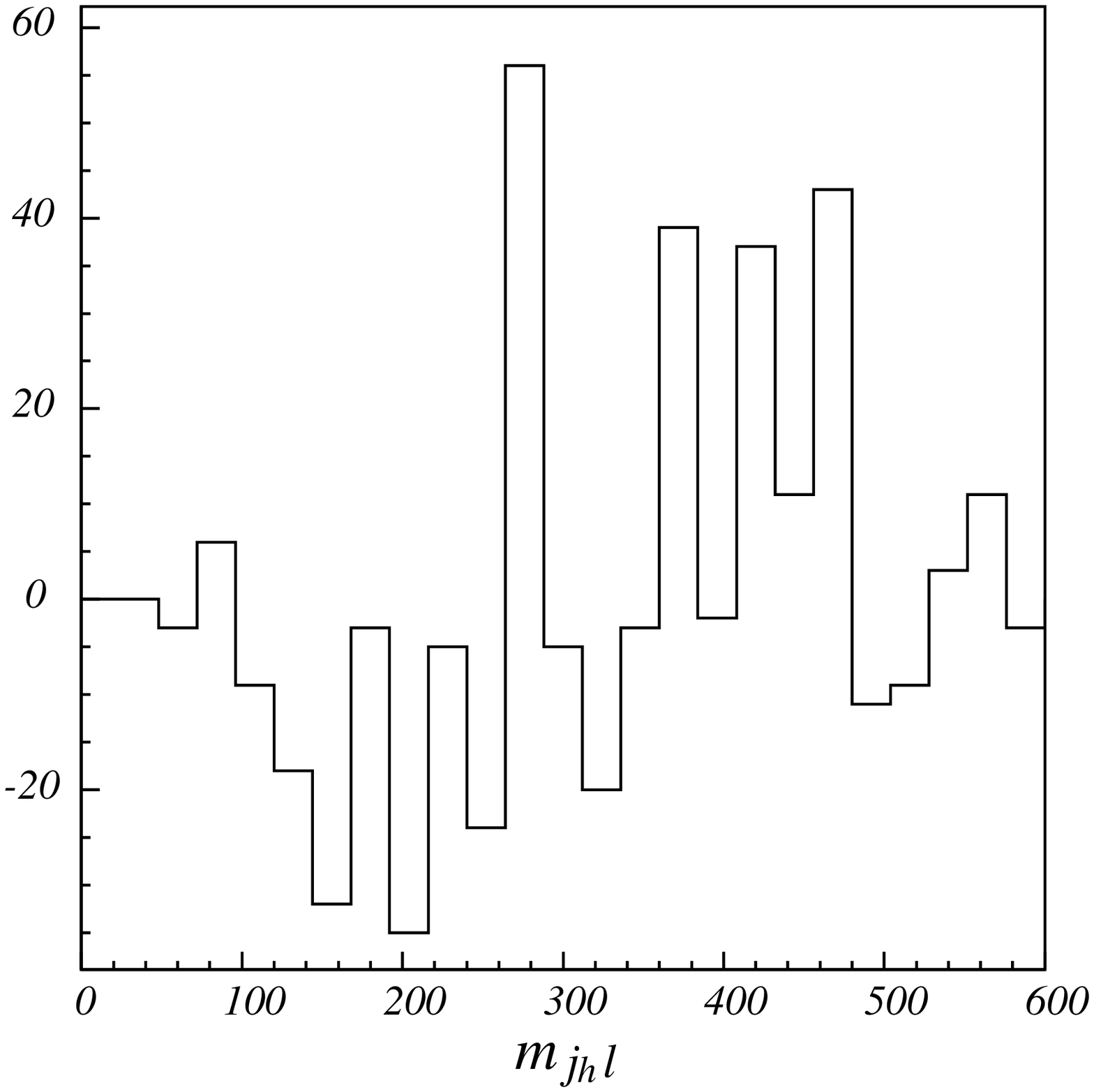}
\caption{The difference between $m_{j_hl^+}$ and $m_{j_hl^-}$ distributions as a function of $m_{j_hl}$ for points A and A2.}
\label{charge_asym}
}

In our study, once we know that the LSP is Bino-like by $m_{T2}$ method and that $\sql$ cascades via long decay chains, this method can be used to probe the property of $\zii$. Figure \ref{charge_asym} shows $N(j_hl^+)-N(j_hl^-)$ as a function of $m_{j_hl}$ for points A and A2 at $\sim 60~fb^{-1}$ of integrated luminosity. In these plots, we use only 
$j_h$, one of the two highest $p_T$ jets which gives higher $m_{jll}$ value. Moreover,
$m_{j_hl}$ is calculated only for lepton with higher $p_T$ because $\mzii-\mer>\mer-\mlsp$ and lepton from $\zii$ decays tends to have larger $p_T$. In addition, we require $m_{ll}<70.7~\GEV$ and $m_{jll}<m_{jll}^{max}$. For point A (left), the charge asymmetry is clearly positive which indicates the $\sql\rar\zii\rar\slr$ decay. On the right panel, the asymmetry is also positive but less prominent. At $m_{j_hl}>300~\GEV$, the distribution is  relatively flat compared to the left panel. This can be interpreted as the contribution from $\sqr$ decay.
The actual ratio of $Br(\sqr\rar\zii)/Br(\sql\rar\zii)$ is 
\beq
	\frac{Br(\sqr\rar\zii)}{Br(\sql\rar\zii)}=\left\{\begin{array}{l}
	0.16\hspace{2cm}\textrm{point A}\\
        0.67\hspace{2cm}\textrm{point A2}\end{array}~~~.\right.\non
\eeq
The ratios differ by factor of four: a factor two from the reduction of $Br(\sql\rar\zii)$ and another from the enhancement of $Br(\sqr\rar\zii)$ (see Table \ref{br_s1s2}). This makes the contribution from $\sqr$ for point A2 is comparable to $\sql$'s contribution and really cause the decrease in the charge asymmetry. Therefore the flatness of charge asymmetry implies that $\zii$ have smaller Wino component and it is persuasive to conclude that $M_1<\mu<M_2$ for this point.

In Figure \ref{ratio_br}, we show the dependence of $R\equiv Br(\sur\rar\zii)/Br(\sul\rar\zii)$ on $\mu/M_1$ and $M_2/M_1$. Here, we took parameter point A and allow $\mu$ and $M_2$ to deviate within 20 percent. The ratio $R$ shows little dependence on $M_2/M_1$ while it ranges from below 0.1 to 0.6 when $\mu/M_1$ ratio varies from 1.5 to 2.7. Due to the form of neutralino mass matrix, Wino mixes with Higgsinos
easier than with Bino. Therefore, as a probe into the reduction of Wino component of $\zii$, the ratio $R$ depends on $\mu/M_1$ stronger than on $M_2/M_1$. When $\mu/M_1$ is smaller, mixing between neutralino states is larger and the charge asymmetry receives more $\sqr$ contribution so that the distribution becomes flat.

\FIGURE[!ht]{
\includegraphics[scale=0.7]{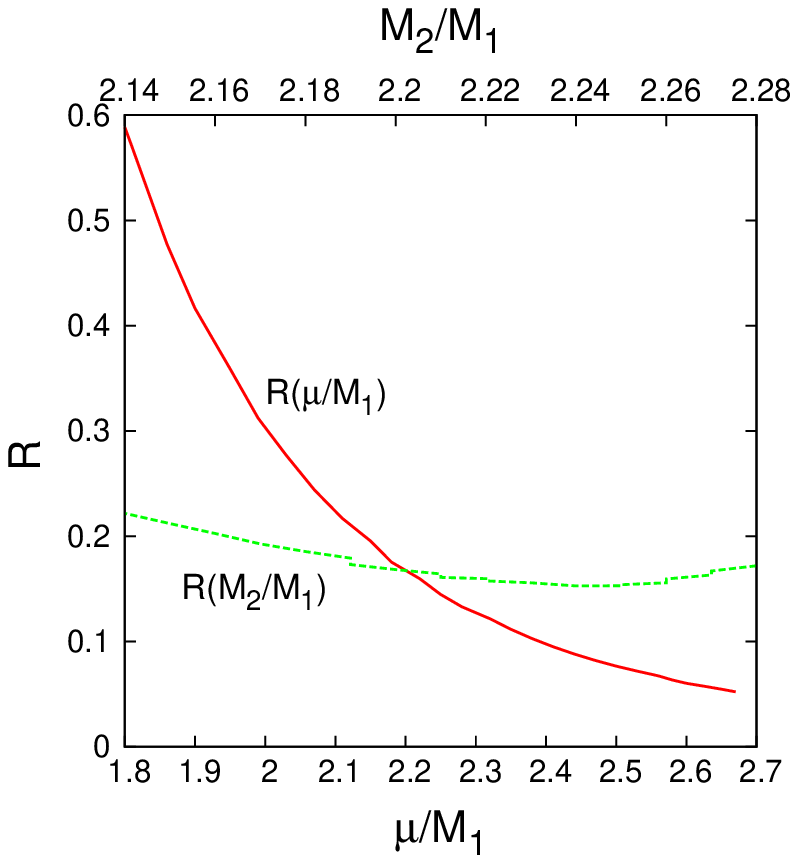}
\caption{The ratio $Br(\sur\rar\zii)/Br(\sul\rar\zii)$}
\label{ratio_br}
}

Now we give comments on other possible flipping solutions. The first possibility is a solution when $M_1<M_2,\mu$ and $\mer$ and $\mel$ are flipped. This case can be easily distinguished by utilizing a charge asymmetry to confirm the chirality of a daughter slepton in the decay $\sql\rar\tilde{\chi}^0_i\rar\tilde{l}_\beta$. However, if not only $\mer$ and $\mel$, but also $\msqr$ and $\msql$, and $M_1$ and $M_2$ are inverted with $M_2<M_1,\mu$, events near $m_{ll}^{max}\sim 71~\GEV$ comes from $\sqr\rar\zii\rar\sll\rar\lsp$ instead and   charge asymmetry of the original and this solutions must be the same.
Nevertheless, whenever $M_1$ and $M_2$ are flipped, the chargino sector will be affected. For example, for $M_2<M_1<\mu$, mass splitting $\mwii-\mwi$ will be close to $\mziv-\mlsp$ and hence endpoint position from ($\wii\rar\snu\rar\wi$) will generally differ from $m_{ll}^{max}\sim 112~\GEV$.

We end this section by a table summarizing possible flipping solutions and methods to resolve it, Table \ref{all_sol}.

\TABLE[!ht]{
\begin{tabular}{|c|c|}
\hline
~Parameters~&~Resolving Methods~
\\ \hline
\hline
$\mu < M_1,M_2$ & $m^{max}_{ll}(\wii\rar\snu\rar\wi)$ + $m_{T2}$
\\\hline
$M_2<M_1$ & $m^{max}_{ll}(\wii\rar\snu\rar\wi)$
\\\hline
$M_1<\mu<M_2$ & branching ratio + charge asymmetry 
\\\hline
$\mel\leftrightarrow\mer$ & charge asymmetry 
\\ \hline
\end{tabular}
\caption{\footnotesize Possible flipping solutions and resolving methods.}
\label{all_sol}
}

\section{Implication to LFV Processes}

We begin this subsection by showing the dependence of $Br(\meg)$ on $\mu$ parameter for three model points, together with current and future experimental bounds, in Figure \ref{cancellation}. In these plots, we took $\mu, M_1$ and $M_2$ of our model points A, A2 and A3, and assumed flavor-violating  parameters as the model with horizontal symmetry described in Section II with $x=0.30$. Here, we ignore the left-handed slepton mixings since we are interested in the cancellation among diagrams.

\FIGURE[!ht]{
\hspace{-.75cm}
\includegraphics[scale=0.5]{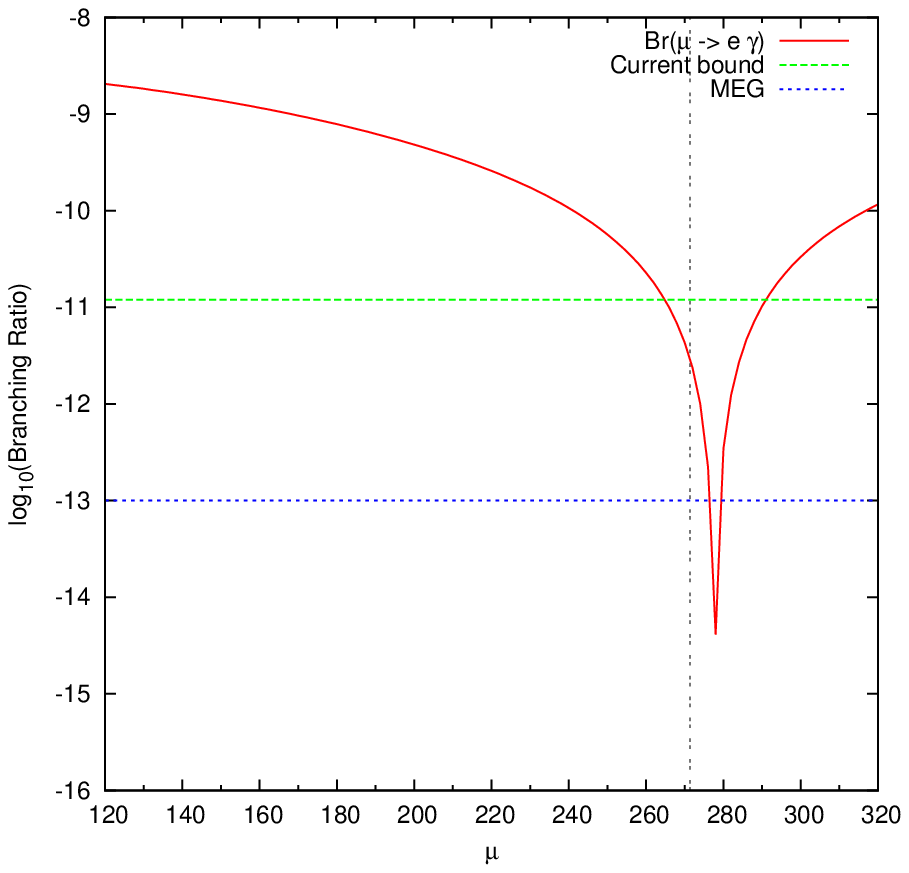}
\hspace{-2cm}
\includegraphics[scale=0.5]{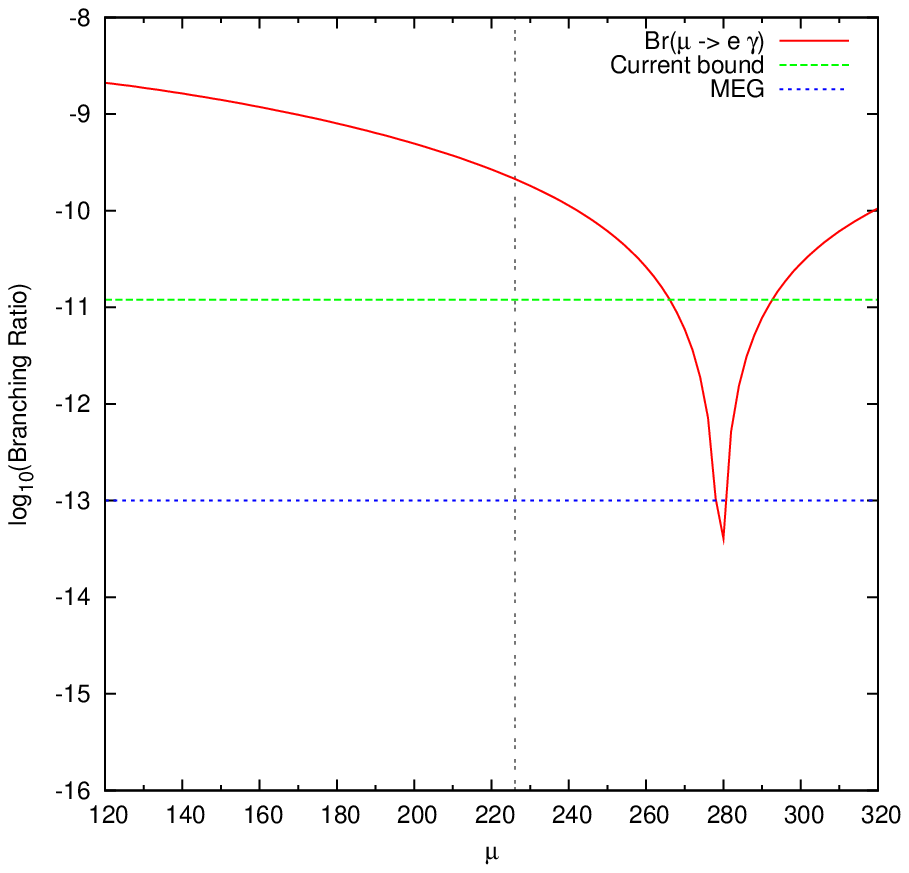}
\hspace{-2cm}
\includegraphics[scale=0.5]{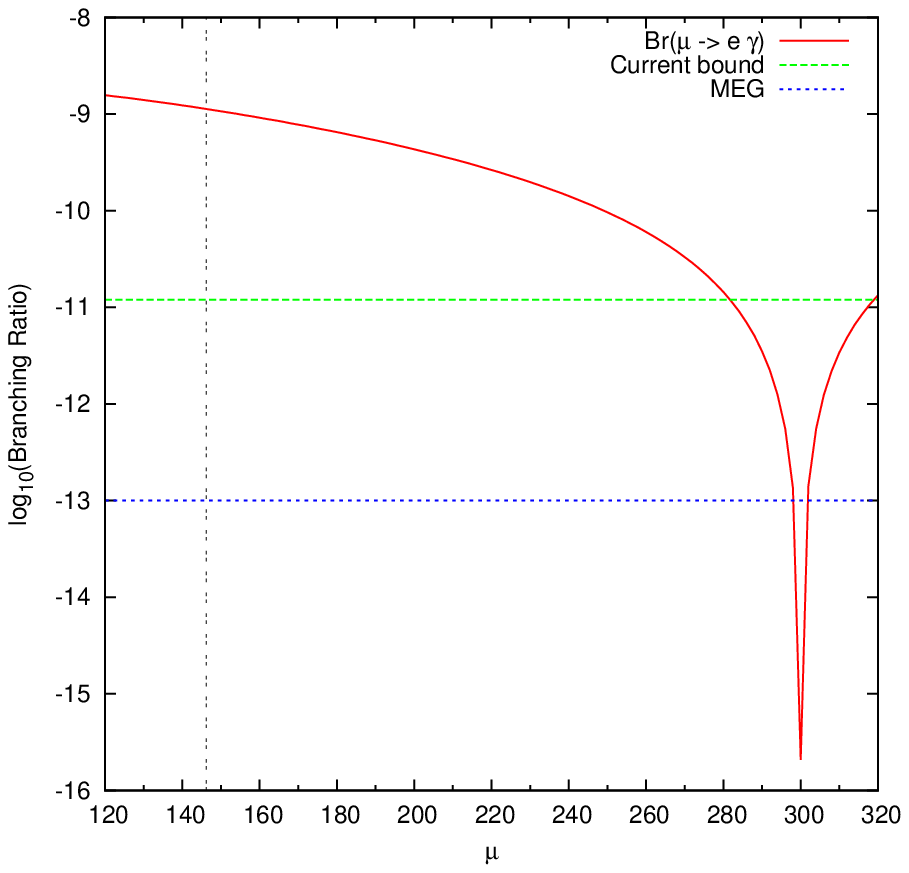}
\caption{$Br(\meg)$ as a function of $\mu$ for points A, A2 and A3 respectively. We assume the MSSM with horizontal symmetry given in Section II.
In these  plots, $x=0.30$ and a vertical line shows the nominal $\mu$ value for each point.}
\label{cancellation}
}

In each plot, there exists a region where the branching ratio becomes very small.\footnote{The difference of the minimum value of the branching ratio in each plot 
is merely the numerical artifact. } This is due to the cancellation explained in Section II. The positions of cancellation point are fairly close for all plots due to the degenerate slepton masses, but only point A is in the cancellation region and then gives $Br(\meg)$ below the current experimental upper bound. Precise determination of $\mu$ parameter is then important for determination of flavor mixing parameter, especially for points A and A2 which have similar collider signature but $Br(\meg)$ differ by two orders of magnitude. In Figure \ref{conv} we show $Br(\meg)$, $Br(\mu\rightarrow eee)$ and 
$\mu\to e$ conversion rate in Titanium ($Br(\mu\rightarrow e; {\rm Ti})$)
as functions of $\mu$ for point A. 
The $\mu\rightarrow e$ conversion rate from the $Br(\meg)$  also 
shows strong sensitivity to the  SUSY parameters at different $\mu$ values. 
Because the cancellation takes place in different region, it is important to measure 
LFV in multiple channels so that we do not miss it accidentally.

\FIGURE[!ht]{
\includegraphics[scale=0.35]{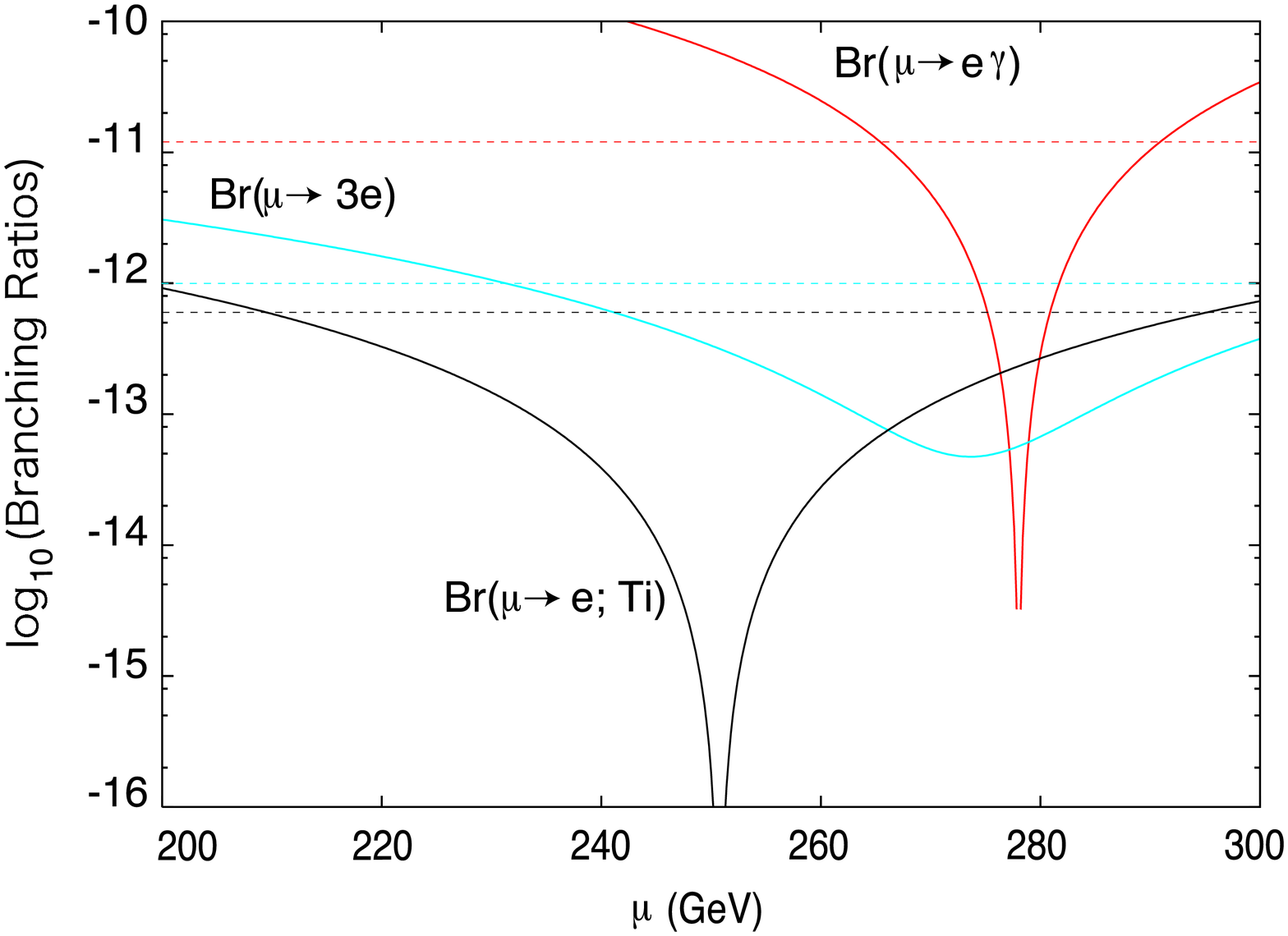}
\caption{ $Br(\meg)$ (red), $Br(\mu\rightarrow 3e)$ (blue) and
  $Br(\mu\rightarrow e; {\rm Ti})$ (black) as functions of $\mu$ for
  point A. Here, dashed lines are for the experimental bounds.}\label{conv}
}

We now estimate the error of the SUSY parameters using information from Table \ref{endpfit}. First of all, we fixed $\tan\beta =10$ and required that $\mzii - \mlsp$, $\mziv - \mlsp$ and $\mlsp$ should not differ from their nominal values greater than 1.5 $\GEV$, 2 $\GEV$ and 7 $\GEV$ respectively. We then looked for the allowed values of $M_1, M_2$ and $\mu$ near their nominal values. Furthermore, we calculated $\mer$ and $\mel$ using $m_{ll (1)}^{max}$ and $m_{ll (5)}^{max}$ information and then calculated $\Delta\chi^2$ defined in  Eq.~(\ref{deltachi2}) from other calculated endpoints, i.e. $m_{jll}^{max}, m_{jl}^{max}, m_{jl}^{min}, m_{ll (3)}^{max}$ and $m_{ll (4)}^{max}$. The central values of $\mu, M_1$ and $M_2$ and their ratios are given with their estimated statistical errors within $\Delta\chi^2=1$ (1-$\sigma$) region in Table \ref{par_constrain}. The absolute values receive sizable errors while the the ratios gives a better sensitivity due to correlation among parameters.  Comparing Figures \ref{cancellation} and \ref{conv}, we can see that  the size of error of $\mu$ is small enough to resolve the position of the cancellation in $Br(\mu \rightarrow e \gamma)$.
\footnote{Since information of $\mu$ is obtained mainly from $\mziv$, it would not be so sensitive to $\tan\beta$. In addition, the cancellation point of $Br(\meg)$ , Eq.~(\ref{cancellation_cond}) is insensitive to $\tan\beta$ neither.}

\TABLE[!ht]{
\begin{tabular}{|c|c|c|}
\hline
~Parameter~&~Central value~ & ~Estimated error~
\\ \hline
\hline
$\mu$ & 271.33 & ${}^{+6.89}_{-6.81}$
\\\hline
$M_1$ & 122.49 & ${}^{+7.16}_{-7.17}$
\\ \hline
$M_2$ & 230.89 & ${}^{+6.57}_{-6.54}$
\\ \hline
\hline
$\mu / M_1$ & 2.215 & ${}^{+0.084}_{-0.078}$
\\ \hline
$M_2 / M_1$ & 1.885 & ${}^{+0.063}_{-0.057}$
\\ \hline
\end{tabular}
\caption{\footnotesize Central values and 1-$\sigma$ constrained intervals of $\mu, M_1$ and $M_2$ and their ratios after constraints from various cascade decay endpoints are imposed as described in the text.}\label{par_constrain}
}

Even if $\mu$ is in the region where $Br(\meg)$  is canceled, the direct slepton mass 
measurement at the LHC would provide the information on the non-universality in the slepton mass matrices. 
As had been emphasized earlier, $\mer$ will be measured precisely. Then if $\mer$ and $\mmur$ are different due to the flavor violating effect, we will measure each of them rather accurately. In addition, the error in $m_{\tilde{e},\tilde{\mu}}-\mlsp$ in Table \ref{invt} tends to cancel when we take the difference $m_{\tilde{e}_{R(L)}}-m_{\tilde{\mu}_{R(L)}}$. We therefore estimate errors of slepton mass difference in Table \ref{deltam} by assuming $\mer-\mmur = 1~\GEV$ and $\mel-\mmul = 2~\GEV$.

\TABLE[!ht]{
\begin{tabular}{|c|c|c|}
\hline
Mass difference& Central value & Estimated error 
\\ \hline
\hline
$\mer - \mmur$ & 1.00 & ${}^{+0.04}_{-0.04}$
\\ \hline
$\mel -\mmul$ & 2.00 & ${}^{+0.48}_{-0.49}$
\\ \hline
\end{tabular}
\caption{\footnotesize One-$\sigma$ error estimation of mass splitting between selectron and smuon in $\GEV$ when $\mer-\mmur = 1~\GEV$ and $\mel-\mmul = 2~\GEV$ are assumed.}\label{deltam}
}

By taking $\mu, M_1$ and $M_2$ of our model point A and assuming flavor-violating  parameters as the model with horizontal symmetry described in Section II, mass difference $\mer-\mmur$ and $Br(\meg)$ are plotted as a function of parameter $x$ in Figure \ref{feng21}. Note that since LFV masses in the left-handed slepton sector are highly suppressed in this model, then left-handed smuon-selectron mass splitting is predicted to be undetectable for this case. The current experimental bound by MEGA \cite{mega} and the MEG sensitivity \cite{meg} are also displayed. The mass difference of order one percent, corresponding to $1\sim 2~\GEV$, is allowed by MEGA bound due to the partial cancellation. 
Even though the statistical error of $\mer-\mmur$ in Table \ref{deltam} are very good, however from the fact that slepton decay width is around $0.2~\GEV$, we simply expect that the LHC has roughly equal sensitivity to the MEG experiment.

\FIGURE[!ht]{
\includegraphics[width=10.4cm]{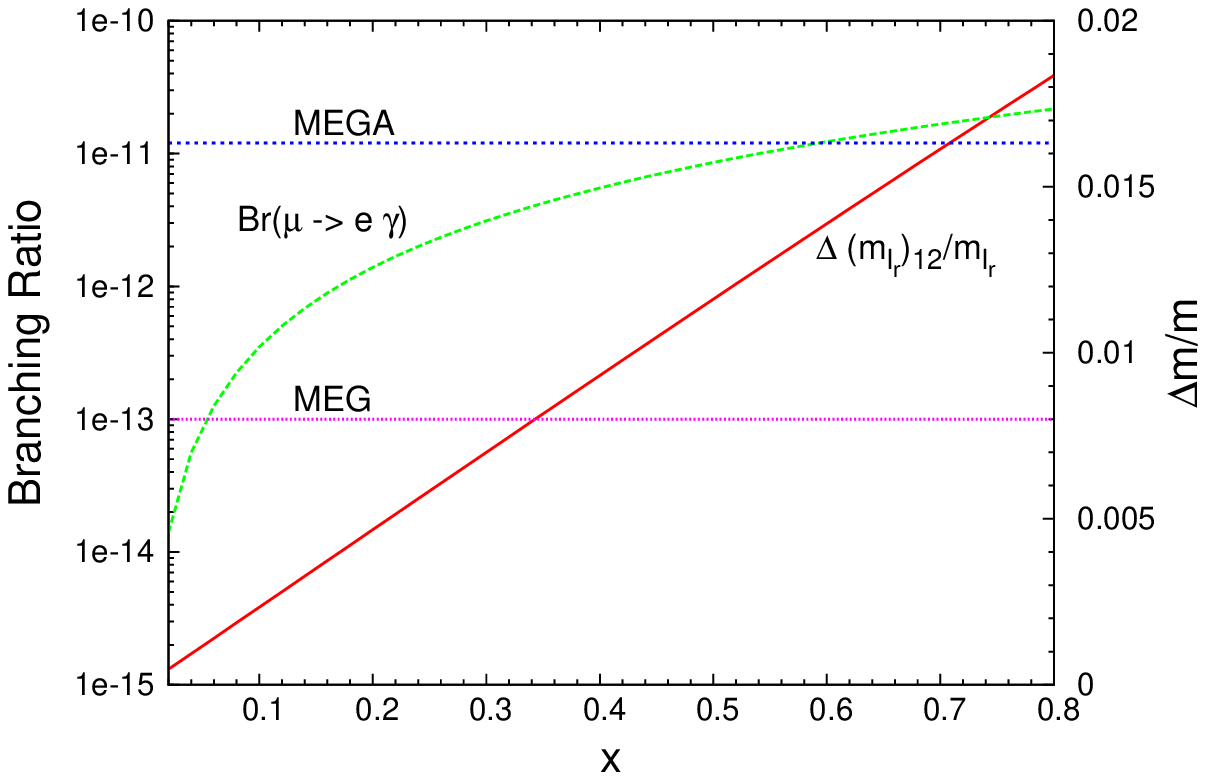}
\caption{Mass splitting $\Delta (\mslr)_{12}/\mslr\equiv 2(\mer-\mmur)/(\mer +\mmur)$ and $Br(\meg)$ in the model with horizontal symmetry as a function of parameter $x$. 
 The current experimental bound and the MEG sensitivity  for $\meg$ are displayed by horizontal lines.}\label{feng21}
}

\FIGURE[!ht]{
\includegraphics[width=10cm]{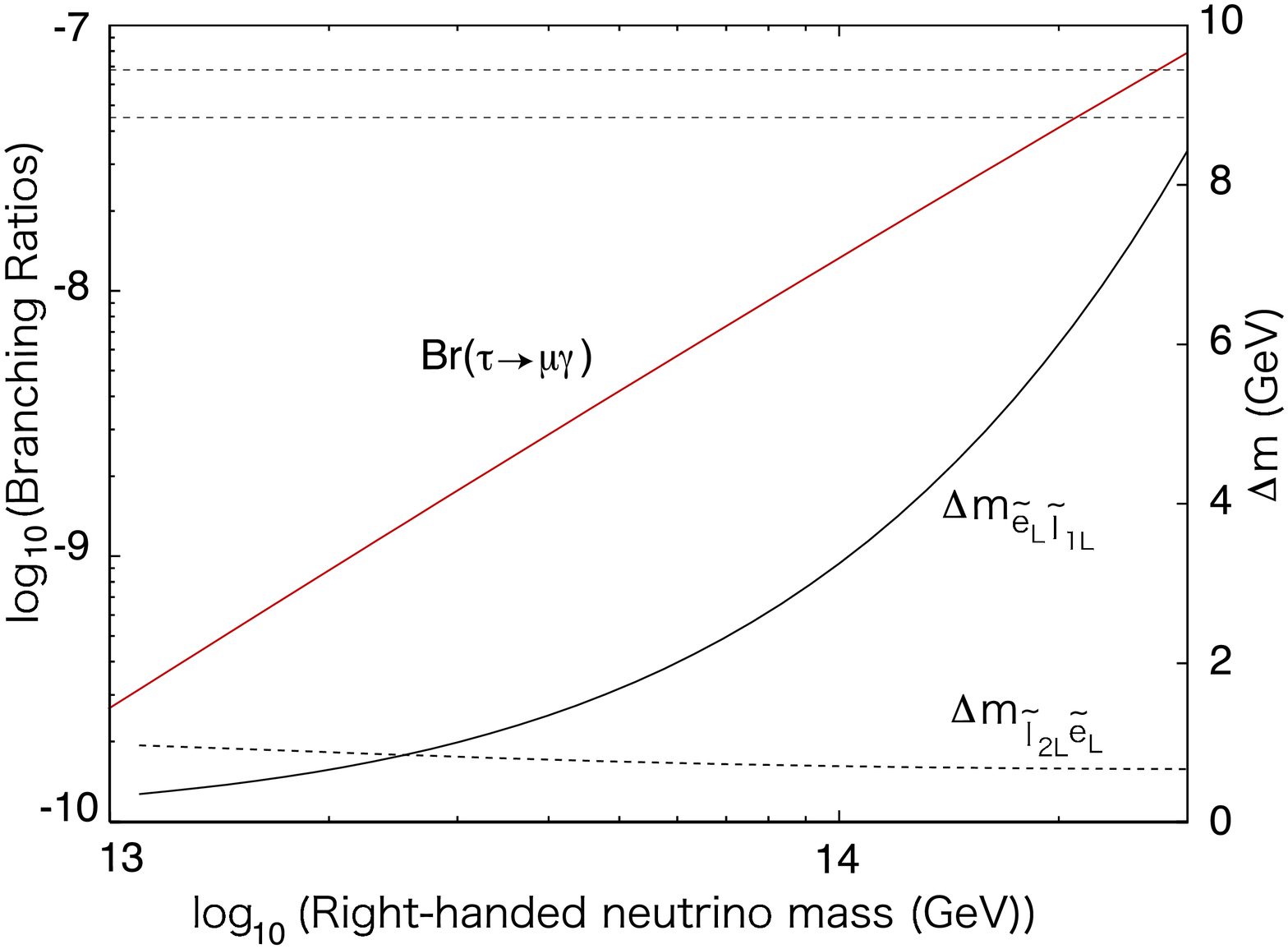}
\caption{Mass splittings, $(\mel-m_{\tilde{l}_{1L}})$ and $(m_{\tilde{l}_{2L}}-\mel)$, and $Br(\tau\rightarrow \mu \gamma)$ in the MSSM with right-handed neutrinos as a function of common right-handed neutrino scale.
${\tilde{l}_{1/2L}}$ is the lighter/heavier mixed state of left-handed smuon and stau. Dashed lines are BABAR and Belle bounds on $Br(\tau\rightarrow \mu \gamma)$.
}\label{nr}
}

 A similar plot for the MSSM with right-handed neutrinos is shown in Figure \ref{nr}. In this model, the left-handed smuon and stau are mixed and form the lighter/heavier state ${\tilde{l}_{1/2L}}$.
 The horizontal axis now represents common right-handed neutrino mass scale. We show $Br(\tau\rar\mu\gamma)$ because it gives the most stringent bound in this model. 
The experimental bounds of BABAR and Belle \cite{Banerjee} allow mass splitting between left-handed selectron and ${\tilde{l}_{1L}}$ of order few $\GEV$.

As mentioned before, point A is in the cancellation region and simultaneously provides an acceptable dark matter relic density. This can be understood by the followings. In Figure \ref{cancellation_spectrum}, we plotted values of $\mu, \mer, \mel, \mzii,$ and $\mziv$ as functions of $m_0$ when $m_{1/2}$ is fixed at $300~\GEV$ and $\tan\beta=10$. In the plot, we assume the mSUGRA relation among slepton and gaugino masses:
\beq
	M_1 &=& 0.4 m_{1/2}, \hspace{1cm} M_2 = 0.8 m_{1/2},\\
	m^2_{\tilde l_L} &=& 
		m_0^2 + 0.5 m_{1/2}^2 -\left(\frac{1}{2} - \sin^2\theta_W \right)m_Z^2\cos 2\beta,\\
	m^2_{\tilde l_R} &=& 		
		m_0^2 + 0.15 m_{1/2}^2 - \sin^2\theta_Wm_Z^2\cos 2\beta,
\eeq
and the value of $\mu$ is obtained from the cancellation condition
shown in Eq.~(\ref{cancellation_cond}). The condition requires $\mu\sim\msll$. Therefore as $m_0$ increases, $\mu$ increases as well. The DM density also increases because Higgsino component of LSP is reduced and scalar masses are increased simultaneously. 
When $m_0$ reaches value around $150~\GEV$, $\Omega_{\rm DM}h^2\sim 0.37$.
In addition, the decay $\zii\rar\ser$ is no longer open and the slepton information may be lost. On the other hand, in a small $\mu$ region, DM relic abundance is small and $\ziv\rar\sel$ and $\zii\rar\ser$ are always kinematically allowed. In the case, the possibility to access to both right- and left-handed slepton masses at the LHC is opened if their couplings are not so small. In the same plot, a cross mark represents actual $\mu$ value for mSUGRA point at $m_0=100~\GEV$. Besides having too large DM abundance, it is very far from the cancellation point for $Br(\meg)$. By reducing $\mu$ from mSUGRA scenario, the acceptable relic density and $\meg$ cancellation conditions can be met concurrently. 

\FIGURE[!ht]{
\includegraphics[scale=0.8]{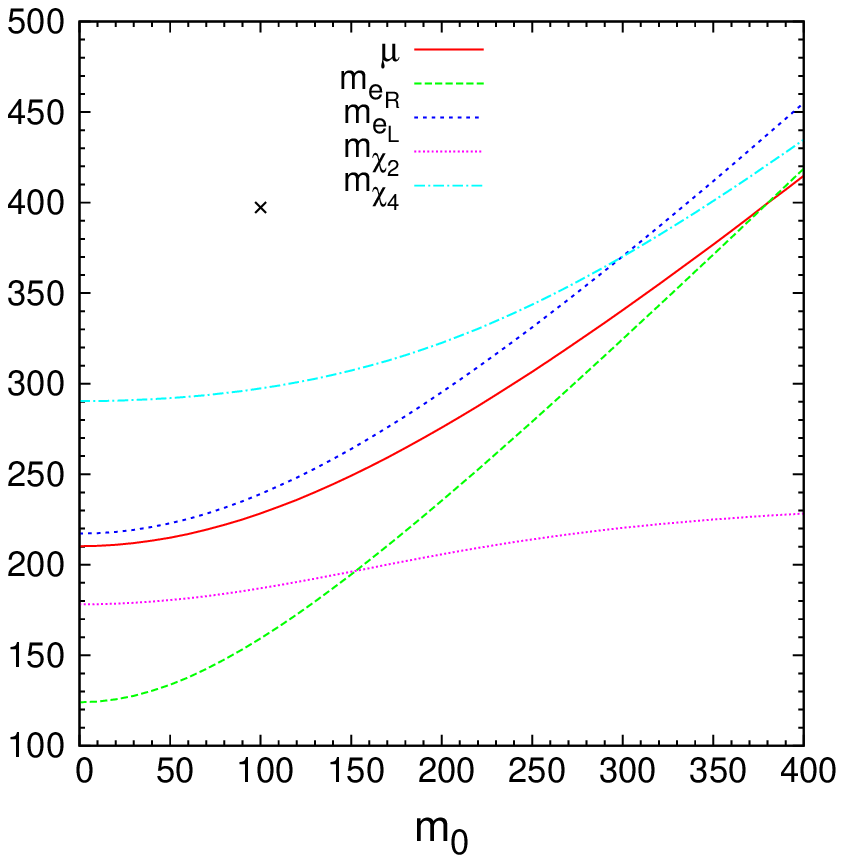}
\caption{Various sparticle masses and $\mu$ value as functions of $m_0$ at fixed $m_{1/2}=300~\GEV$. A cross mark represents $\mu$ value of mSUGRA point at $m_0=100~\GEV$.}\label{cancellation_spectrum}
}
%

\TABLE[!ht]{
\begin{tabular}{|c|c|c|}
\hline
~&~$\Omega_{\textrm{DM}}h^2$~&
~$\sigma^{SI}_{p\chi} (10^{-8}~pb)~$\\ \hline
\hline
~~A~~~& 0.1179 & 1.55
\\ \hline
~~A2~~ & 0.0817 & 3.15 
\\ \hline
~~A3~~ & 0.0096 & 17.50 
\\ \hline
\end{tabular}
\caption{\footnotesize Dark matter relic density and spin independent LSP-proton scattering cross section for points A, A2 and A3.} \label{omega_a2a3}
}

Finally we comment on the dark matter direct detection cross section of our model points. The DM density and spin independent LSP-proton scattering cross section, $\sigma^{SI}_{p\chi}$, are evaluated for points A, A2 and A3 (Table \ref{omega_a2a3}).  Points A2 and A3 are ruled out by relic abundance if only one species of DM is assumed. Moreover, Point A3 seems to be ruled out already by the recent direct detection bound from CDMS, $\sigma^{SI}_{p\chi} < 4.6\times 10^{-8}~pb$ \cite{cdms}. However, uncertainty in $\sigma^{SI}_{p\chi}$ is large due to three important sources. The first one comes from nucleon matrix element of strange quark, i.e. $\langle p|m_s{\bar s} s|p\rangle$, whose recent calculation was found to be an order of magnitude smaller than the previous ones \cite{Ohki}. Another two sources are $\tan\beta$ and pseudo-scalar Higgs mass, $m_A$. The pseudo-scalar Higgs exchange diagram gives important contribution proportional to $\left(\frac{\tan^2\beta}{m_A^4}\right)$. In our analysis, we just take $\tan\beta=10$ and do not fix $m_A$. 
However, if the pseudo-scalar Higgs is heavier or $\tan\beta$ has a smaller value than our sample point, the $\sigma^{SI}_{p\chi}$ would become smaller and evade the direct detection bound.

\TABLE[!ht]{
\begin{tabular}{|c|c|c|}
\hline
~$\mu/M_1$~&~$\Omega_{\textrm{DM}}h^2$~&
~$\sigma^{SI}_{p\chi} (10^{-8}~pb)~$\\ \hline
\hline
1-$\sigma$ & 0.1101 - 0.1271 & 1.37 - 1.76 
\\ \hline
3-$\sigma$ & 0.0929 - 0.1498 & 1.02 - 2.33 
\\ \hline
\end{tabular}
\caption{\footnotesize Dark matter relic density and spin independent LSP-proton scattering cross section within 1-$\sigma$ and 3-$\sigma$ deviations of the $\mu / M_1$ from its nominal value. Here, we assume point A.} \label{omega}
}

We also show in Table \ref{omega} the sensitivity of $\Omega_{\textrm{DM}}h^2$ and $\sigma^{SI}_{p\chi}$ to $\mu/M_1$ ratio for point A. We assumed NUHM with $\mhu\ne m_0$ and changed $\mhu$ value so that $\mu/M_1$ deviates from its central
value by 1-$\sigma$ and 3-$\sigma$ respectively. The uncertainty of $\sigma^{SI}_{p\chi}$ is larger than that of $\Omega_{\rm DM}h^2$ unless we fix $m_A$ and $\tan\beta$. Although we assume rather good determination of $\mu/M_1$ ratio, the uncertainty in $\sigma^{SI}_{p\chi}$ is not small, i.e. nearly factor 1.5 for $1$-$\sigma$.

\section{Discussion}

In this paper, we studied LHC signature of the one-parameter-extended NUHM,   $\mhu\ne\mhd=m_0$. The choice of the boundary condition allows the low energy mass spectrum $M_1<M_2<\mu\sim m_{1/2}$. In that case,  the LSP relic density may be 
 consistent with cosmological and astrophysical observations because pair-annihilation cross section of the lightest neutralino will be enhanced by the Higgsino components. 
 We are especially interested in the region where cancellation among leading contributions to $Br(\meg)$ occurs in the models with right-handed LFV masses because the prediction of $Br(\meg)$ depends strongly on the EW parameters. This cancellation occurs in $\msl \sim \mu$ region,  therefore, we take  a model point with $\mlsp < \mer < \mzii < \mel < \mziv$ as an example. Both the left- and right-handed sleptons can be directly produced via neutralino decays at the model point.  

We  investigated how well SUSY parameters can be determined at the LHC 
for this choice of parameters. In the region when $M_1<M_2<\mu\sim m_{1/2}$ , $\ziv$ and $\wiipm$ are mixed states with rather large Wino components. They are frequently produced from $\sql$ decay and their decay modes into $\sel$ or $\snu$ have large branching ratios. If kinematically allowed,  $\sel$ would decay dominantly into  Wino-like $\zii$; however, $\ziv$ and Wino-like $\zii$ also have small Bino component and they   could decay into $\ser$. Accordingly, various decay patterns shown in  Eq.~(\ref {heavier_decay}), are expectable at the LHC, allowing precise mass determination using endpoint method. 

However, even all above decay could be measured at the LHC and masses of all sparticle involving in the decay chains would be identified,  there are different regions of SUSY parameter space which have mass spectrum consistent with  measured end points. This ambiguity mostly occurs in the neutralino sector. 

For our model point, we find three solutions with similar mass spectrum but with different ordering of $\mu, M_1$ and $M_2$  when the relation $M_1<M_2$ is kept.  These three points predict different $Br(\meg)$  for the same LFV slepton masses. Because the original point is so close to the cancellation point, the prediction could  differ by factor of $\mathcal O$(100) among the three points. The thermal relic density $\Omega_{\rm DM}h^2$ and $\sigma^{SI}_{p\chi}$ also differ by  $\mathcal O$(10). 
However, we find that the reordering leads to different properties of neutralinos which will clearly reflect in their decay branching ratio.
 We showed that $Br(\tilde\chi_i\rar 2l+X)$, rate of events in two hard jets+missing $E_T$ channel, and charge asymmetry play important role to lift the degeneracy.  This is  an excellent example of the complementarity of the LHC and the rare decay searches in SUSY studies.

\section*{Acknowledgments}

This work is supported in part by the World Premier International Center  
Initiative (WPI Program), MEXT, Japan and the Grant-in-Aid for Science  
Research, Japan Society for the Promotion of Science (No.~20244037 and No.~2054252 for JH and No.~16081207 and No.~18340060 for MN) and the Monbukagakusho (Japanese Government) Scholarship (for WS).

\appendix
\section{Appendix}
\subsection{Trilepton and Four-lepton Distribution in Four-lepton Events}

We have shown in Section III that the four-lepton events from $\ziv\rar\sel\rar\zii\rar\ser\rar\lsp$ is a useful mode particularly when $Br(\ziv\rar\sel\rar\zii)$ is sizable but its $m_{ll}$ edge can not be seen in two-lepton events. In this Appendix, we show $m_{lll'}$ and $m_{lll'l'}$ distributions as a consistency check. 

Again, we selected four-lepton events by the cuts:
\begin{itemize}
\item
exactly two pairs of OSSF leptons where each lepton has $p^l_T > 10 ~\GEV$ 
and $|\eta| < 2.5$,
\item
four leptons must be composed of exactly one $e^+e^-$ pair and 
one $\mu^+\mu^-$ pair.
\end{itemize}
 For each event which passes the cuts, we calculated $m_{lll'}$. The cascade decay is expected to have two trilepton endpoints:\\
(1) One from the upper part of the decay chain, i.e.
$\ziv\rar\sel\rar\zii\rar\ser$. 
The expected endpoint $m_{lll'}^{\textrm{max}} = 192.8 ~\GEV$.\\
(2) Another from the lower part of the decay chain, i.e.
$\sel\rar\zii\rar\ser\rar\lsp$. 
The expected endpoint $m_{lll'}^{\textrm{max}} = 123.9 ~\GEV$.

\FIGURE[!ht]{
\includegraphics[scale=0.3]{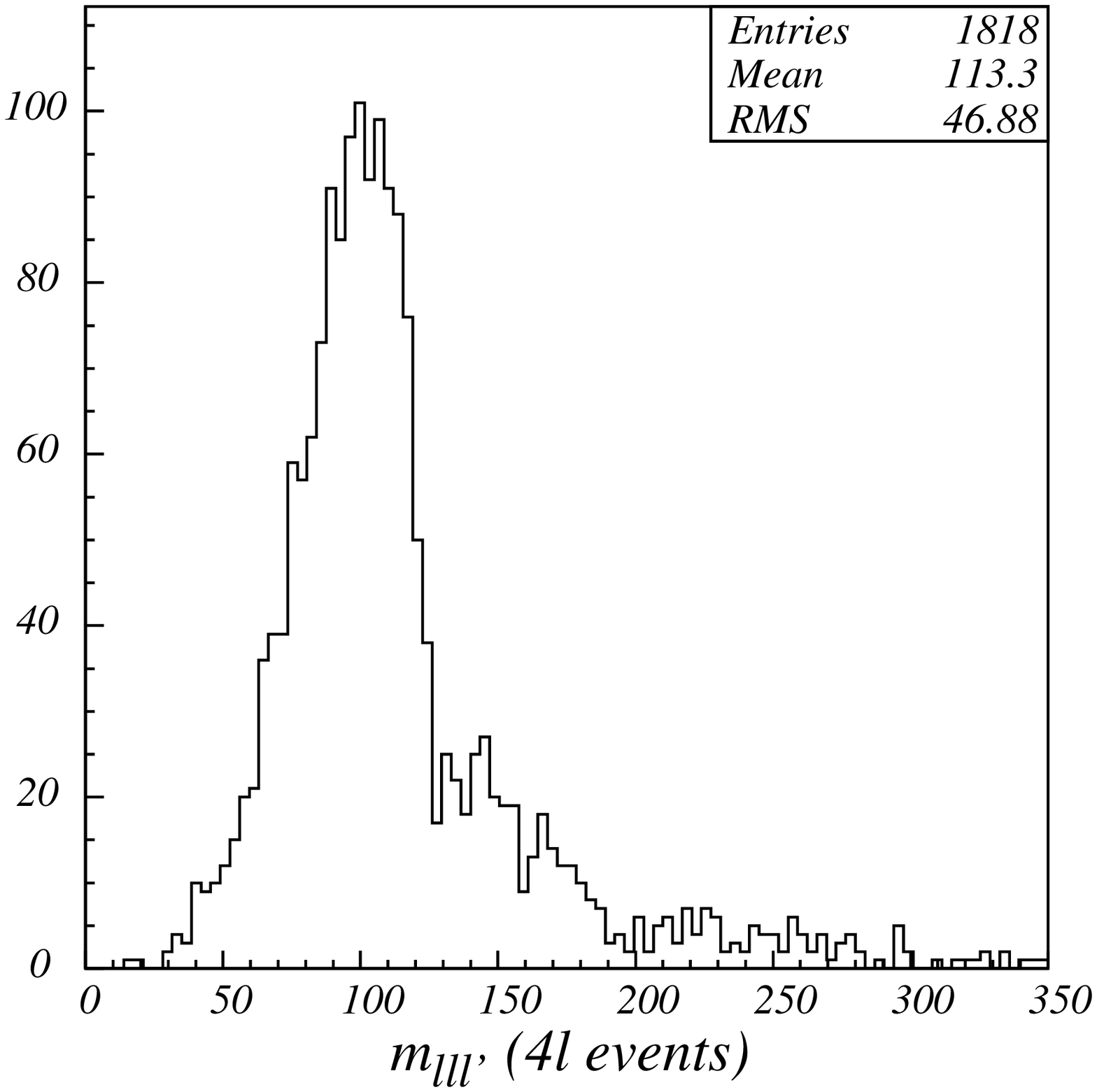}
\caption{The trilepton invariant mass $m_{lll'}$ distribution (in $\GEV$) which satisfies $m_{ll} < 70.7 ~\GEV$ and $71 ~\GEV < m_{l'l'} < 123 ~\GEV$.}
\label{3l}
}

By utilizing the $m_{ll}^{\textrm{max}}\sim 123~\GEV$ from $\ziv\rar\sel\rar\zii$ as in Section III, one can further find additional constrain on sparticle masses by looking at the endpoint of the $m_{lll'}$ distribution as follows.
Principally, \\
(1) if one requires $m_{ll} < 70.7 ~\GEV$ and $71 ~\GEV < m_{l'l'} < 123 ~\GEV$, one expects to see the trilepton endpoint from the lower part of the decay chain, or\\
(2) if one requires $m_{l'l'} < 70.7 ~\GEV$ and $71 ~\GEV < m_{ll} < 123 ~\GEV$ instead, the trilepton endpoint from the upper part of the decay chain is expected to show up.

However, in our Monte Carlo study, the endpoint from the upper decay chain is invisible.  This may be because the distribution for the upper endpoint spreads over wider range so that its height of the peak is lower, and then it is buried under backgrounds. On the contrary, the one from the lower decay chain in Figure \ref{3l} is rather impressive as it may receive contribution from chargino decay.

Finally, $m_{lll'l'}$ distribution is shown in Figure \ref{4l}. The edge is expected not to 
exceed the mass difference $\mziv - \mlsp = 208.5 $. In the left panel, $m_{ll}$ of one OSSF pair is required to be $< 70.7 ~\GEV$ and that of other OSSF pair is requred to be in the range $71 ~\GEV < m_{l'l'} < 123 ~\GEV$. Once we further impose the constrain $m_{lll'} < 124 ~\GEV$, the tail almost disappears as shown in the right panel.

\FIGURE[!ht]{
\includegraphics[scale=0.3]{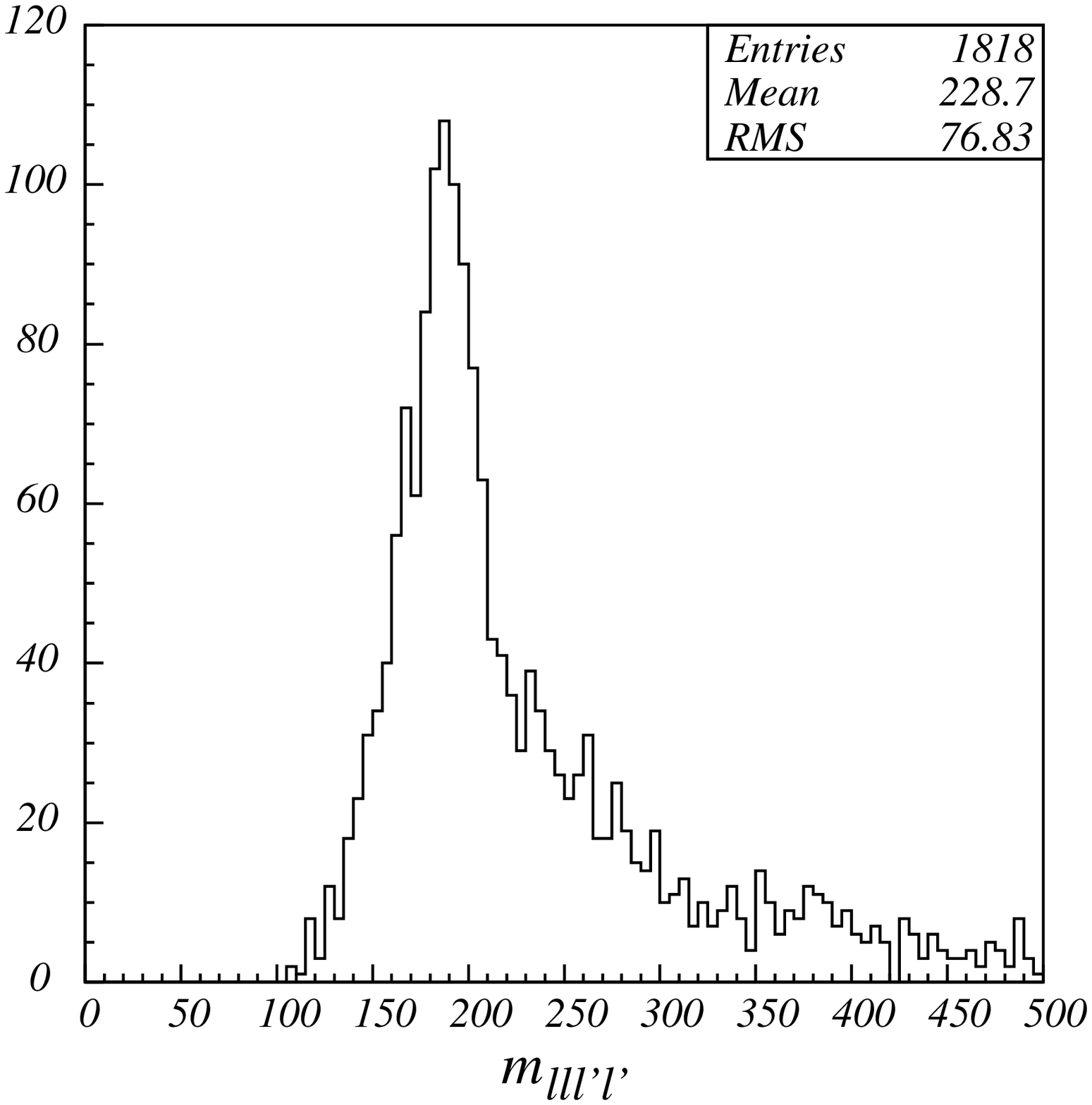}
\includegraphics[scale=0.3]{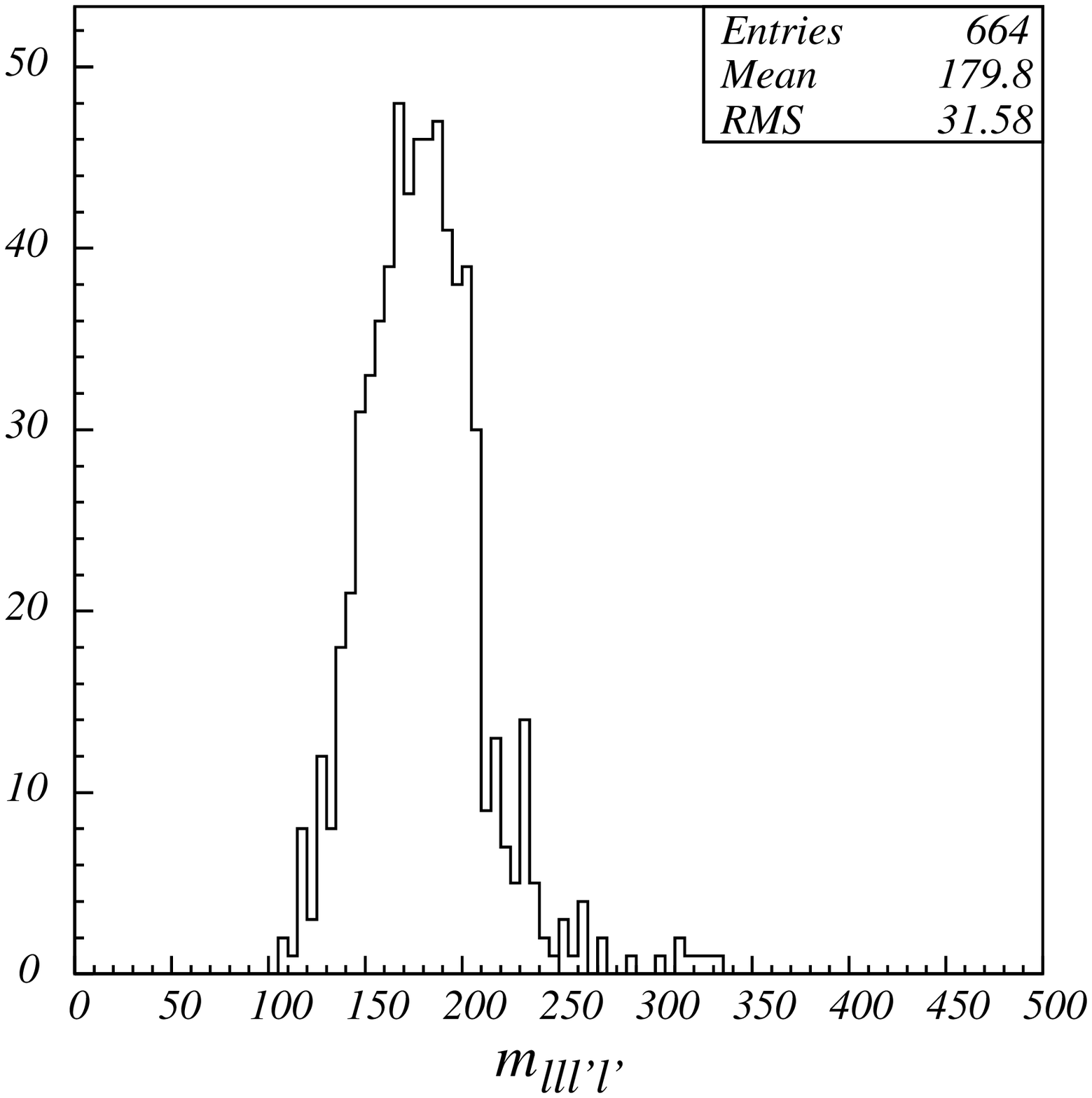}
\caption{The four-lepton invariant mass $m_{lll'l'}$ distribution (in $\GEV$) satisfying $m_{ll} < 70.7 ~\GEV$ and $71 ~\GEV < m_{l'l'} < 123 ~\GEV$ (left panel). The right panel shows the same distribution when $m_{lll'} < 124 ~\GEV$ is further imposed.}
\label{4l}
}


\end{document}